# Superfluorescence as a cooperative amplifier of hidden anisotropy in a ferroelectric hybrid perovskite

Changhao Gao[1]†, Daniel Sandner[1]†*, Talia Ruhrberg Estévez[1], Jeron Timmer[1], Sophia Klubertz[2], Martijn Kemerink[2], Felix Deschler[1*]

[1] Physikalisch-Chemisches Institut, Universität Heidelberg, 69120 Heidelberg, Germany

[2] Institute for Molecular Systems Engineering and Advanced Materials, Universität Heidelberg, 69120 Heidelberg, Germany

† These authors contributed equally to this work

*Corresponding authors, Email: daniel.sandner@uni-heidelberg.de, deschler@uni-heidelberg.de

**Funding:** F.D. acknowledges funding by the German Research Foundation (DFG) via Research Training Group GRK 2948/1.

C.G. acknowledges financial support from the China Scholarship Council (CSC).

Part of this research has been funded by the Deutsche Forschungsgemeinschaft (DFG, German Research Foundation) – project number 545050087.

**Abstract**

Superfluorescence (SF), intense picosecond bursts from self-synchronizing dipoles, is promising for room-temperature quantum sources. Yet, whether this synchronization can convert weak structural anisotropies into robust macroscopic order remains an open question. Here, we demonstrate that polycrystalline ferroelectric 2D perovskite thin films exhibit strongly linearly polarized SF with a degree of polarization (DOP) up to 86%. Remarkably, macroscopic polarization emerges without external bias despite randomly oriented microscopic domains, suggesting a cooperative gain mechanism that amplifies weak local anisotropies by several orders of magnitude into coherent, linearly polarized light bursts. Our results establish cooperative superfluorescence as a general concept for translating microscopic anisotropy into robust macroscopic order, exemplified here for linear polarization, offering a sensitive probe of hidden material symmetries and a framework for polarized quantum light sources.

## Introduction

Superfluorescence (SF) is a cooperative quantum process in which initially incoherent dipoles spontaneously synchronize via quantum fluctuations, producing intense, coherent light bursts (1–4). Macroscopic dipole coherence builds up while competing with dephasing from structural disorder and many-body interactions. Following Dicke's framework (1), collective radiative phenomena have been observed in a variety of condensed-matter systems, for example quantum dots (4, 5), bulk excitonic crystals (6), and magneto-plasmas (7, 8).

Recent work has shown that hybrid metal-halide perovskites generate SF at room temperature despite local disorder and thermal lattice motion, offering a solid-state platform for quantum emitters, collective exciton studies, and symmetry-dependent optical coherence (2, 3, 7, 9–16). This behavior is attributed to a "quantum vibration-isolation" mechanism mediated by large polarons: photoexcited carriers couple to slowly varying lattice distortions that shield electronic dipoles from rapid thermal fluctuations, forming an extended soliton-like state (17). Consequently, the synchronizing dipoles become linked to structural degrees of freedom, suggesting static or dynamic lattice engineering as routes to quantum-state control. Indeed, structural chirality has been used to select the handedness of cooperative emission in perovskite superlattices via photonic chiral spin–orbit coupling (11).

It remains an open question to what extent material anisotropies can compete with quantum fluctuations to seed a directional bias, of any property, in the dipole synchronization — a bias that the collective build-up would then amplify. Such amplification would establish SF as a sensitive probe of material symmetries, and as a strategy for achieving macroscopic photonic order without the high degree of structural control that remains thermodynamically difficult to achieve in solution-processed or polycrystalline thin films.

As a model system, we choose a polycrystalline ferroelectric perovskite. Ferroelectric semiconductors stand out since their persistent intrinsic polarization imposes a well-defined dipole orientation (18, 19). In layered hybrid metal-halide perovskites, ferroelectric polarization can arise from organic cations and distortions of the inorganic framework, while the inorganic slabs retain strong excitonic optical responses and radiative emission (20–22). Such in-plane polarization is shown in Fig. 1A for isopentylammonium-cesium lead bromide.

Ferroelectric polarization implies internal fields and broken symmetry, both of which can modify the local electrostatic environment of photoexcited states, thereby influencing band-edge excitonic transitions, optical anisotropy, and radiative recombination. In this manner, ferroelectric order can be linked to photoluminescence (23–28). These structural and electrostatic anisotropies may also provide a preferred direction for dipole synchronization, raising the question of whether ferroelectric hybrid halide perovskites can impose dipole order in SF. At the same time, in polycrystalline thin films, individual grains and ferroelectric domains are typically much smaller than the excitation spot or the coherently emitting region, so their random orientation averages out to a very weak net optical anisotropy.

Here we report linearly polarized SF from polycrystalline thin films of the ferroelectric quasi-2D perovskite isopentylammonium-cesium lead bromide i-PCPB: $(\text{i-PA})_2\text{Cs}_{n-1}\text{Pb}_n\text{Br}_{3n+1}$ ($\text{i-PA}^+$ = isopentylammonium), despite negligible linear dichroism and a lack of polarized photoluminescence. The emission exhibits hallmark SF signatures: a clear threshold, delayed pulse formation, and temporal narrowing. Remarkably, the SF bursts reach a linear degree of polarization (DOP) of 86%, demonstrating strong polarization amplification despite the films' modest structural anisotropy. Using a modified Dicke model, we show that local dipole anisotropies at the permille level are magnified by several orders of magnitude into macroscopic polarization.

## Observation of superfluorescence in a ferroelectric 2D perovskite

First, we examine the fluence-dependent emission dynamics of thin films of the ferroelectric 2D perovskite *i*-PCPB at room temperature. Here, *i*-PCPB denotes the isopentylammonium-cesium lead bromide quasi-2D perovskite system rather than a single n-phase composition. The homologous layered phases in this material family are described by $(i\text{-PA})_2\text{Cs}_{n-1}\text{Pb}_n\text{Br}_{3n+1}$ with $n = 2 - 4$ and have been reported to be ferroelectric (*20*). The absorption spectrum of the films contains several excitonic features, indicating the coexistence of multiple layered phases (see Fig. S1). As shown schematically in Fig. 1A, *i*-PCPB forms a layered perovskite structure in which organic $i\text{-PA}^+$ cations separate inorganic lead bromide slabs. This reduced-dimensional and ferroelectric lattice provides a structurally anisotropic environment for excitons, making it a suitable platform for investigating how cooperative emission emerges in an anisotropic semiconductor thin film. In the SF process, collective light-matter coupling drives spontaneous dipole synchronization and the formation of macroscopic polarization. Typically, this coherence build-up competes with dephasing and disorder in the solid-state film. Yet, when the cooperative interactions inducing coherence dominate, the macroscopic polarization emerges and decays as a delayed and intense SF burst (Fig. 1B). The fluence-dependent emission spectra for *i*-PCPB (Fig. 1C) reveal the signs of SF. Samples have been excited with 400 nm (3.1 eV) laser pulses in all experiments. We observe a transition from broad spontaneous PL spectra, to a narrow, red-shifted emission band, above a threshold fluence of approximately 62 $\mu\text{J cm}^{-2}$. The onset of this above-threshold emission is further resolved in the time domain in Fig. 1D, where we find this emission to originate from a delayed burst, with time zero calibrated using the prompt below-threshold PL signal (see Fig. S4). With increasing excitation fluence, the burst maximum shifts progressively to shorter delay times, indicating faster build-up of the cooperative macroscopic coherence. Consistently, Fig. 1E shows that the extracted delay time $\tau_D$ decreases from approximately 7.5 to 5.3 ps, while the radiation time $\tau_R$ from 4.9 to 1.1 ps over the measured fluence range, both following the expected and previously reported dependence for cooperative emission indicative of SF (see Section 17 in the SI).

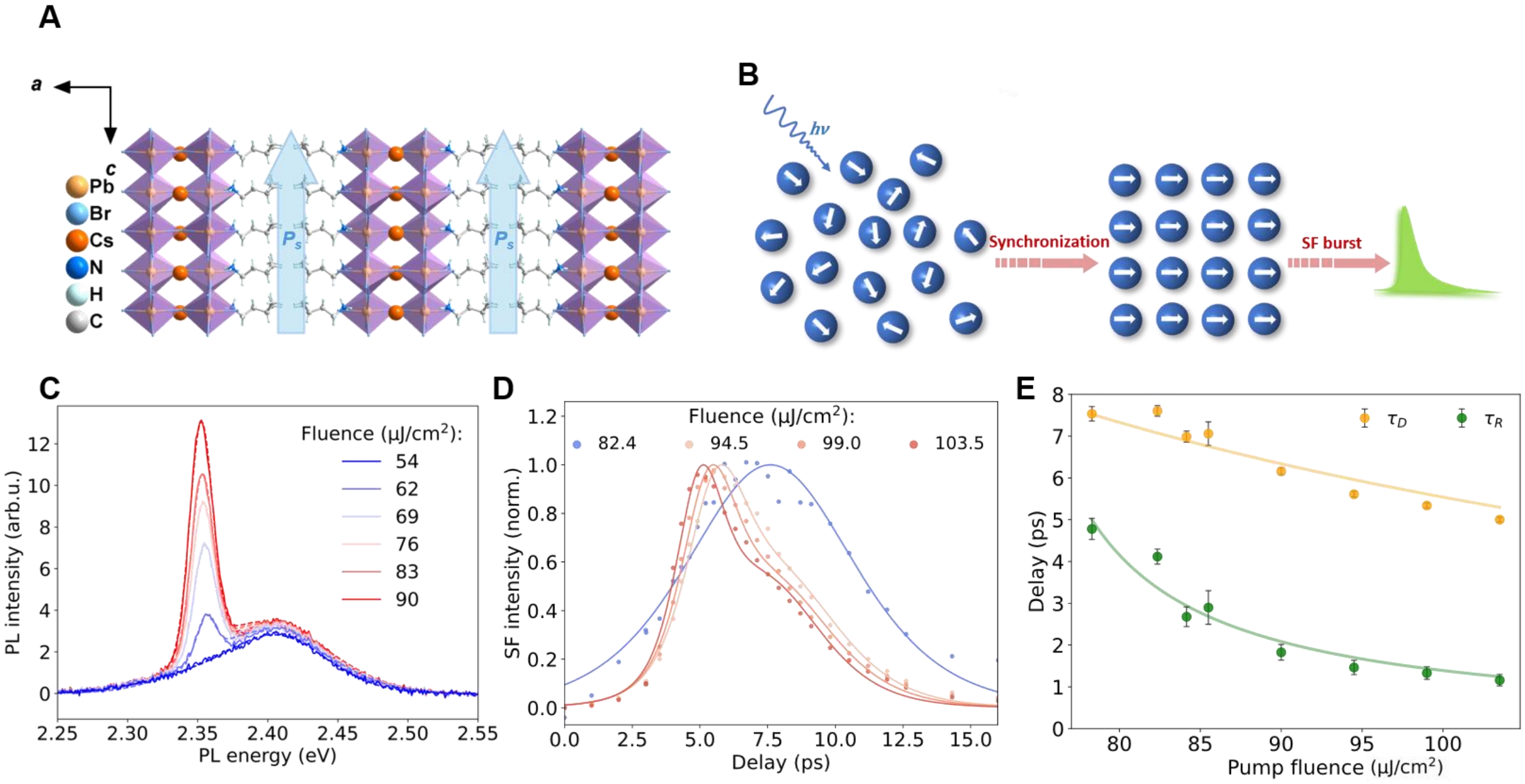

**Fig. 1. Structure and fluence-dependent formation of SF in *i*-PCPB thin films. (A)** Representative (n = 2) Schematic crystal structure of the ferroelectric 2D perovskite *i*-PCPB. **(B)** Illustration of ultrafast excitation followed by spontaneous macroscopic dipole formation and synchronization giving rise to a delayed SF burst. **(C)** Fluence-dependent emission spectra of *i*-PCPB thin film at room temperature showing a narrow, red-shifted SF peak above threshold. **(D)** Normalized time-resolved SF transients at different excitation fluences. **(E)** Fluence-dependent SF delay time $\tau_D$ and burst width $\tau_R$ fitted with SF scaling.

## Strong linearly-polarized superfluorescence in ferroelectric *i*-PCP

The data presented in Fig. 1 shows that the narrow-band emission originates from SF. In the following, we discuss its polarization properties. Figure 2A shows time-averaged emission spectra acquired above the SF threshold with a wire-grid analyzer placed before the spectrometer. While the broad photoluminescence (PL) peak remains unchanged, the intensity of the narrow SF peak is strongly modulated with analyzer orientation. The resulting 180° periodicity is characteristic of linearly polarized emission (see similar curves at 0°, 180° and 90°, 270°).

To determine whether the polarization axis is governed by intrinsic material properties, such as the crystallographic orientation, ferroelectric domains, or strain, rather than by the excitation polarization, we rotated the sample while probing the same excitation spot. To separate the broad PL and narrow SF contributions, each spectrum was fitted with two Gaussian functions (see dashed lines in Fig. 1C and Fig. 2A). Figure 2B shows the SF intensity (dots) as a function of analyzer angle for selected sample orientations. The polar plots are normalized to the minimum and maximum SF intensities. The excellent agreement with Malus' law (solid lines) confirms purely dipolar, linearly polarized emission without measurable higher-order contributions. Notably, the polarization axis rotates together with the sample, demonstrating its intrinsic origin. This suggests that any excitation polarization is rapidly lost following above-gap excitation, likely through ultrafast carrier scattering, before cooperative synchronization sets in. Figure 2C summarizes the polarization axis and degree of polarization (DOP) for each sample orientation.

In contrast, ground-state absorption exhibits negligible linear dichroism without correlation between sample orientation and absorption anisotropy (see Fig. S5). Together with the weak polarization of the PL, these observations indicate that the excitation spot contains many randomly oriented microscopic domains whose anisotropy is averaged out in conventional optical spectroscopy. Remarkably, the spontaneously synchronized macroscopic dipole formed during SF is sensitive to the residual local anisotropies within the excitation spot and cooperatively amplifies them during the synchronization process, giving rise to a robust macroscopic linearly polarized emission state in superfluorescence.

The intrinsic polarization of SF raises the question of when it emerges during the spontaneous synchronization of dipoles that precedes the emission burst. To address this, we performed femtosecond up-conversion spectroscopy with polarization resolution (Fig. 2D). Because sum-frequency generation is intrinsically polarization selective owing to the $\chi^2$ nonlinear interaction and phase-matching conditions, the emitted light was first projected onto the desired linear polarization state and subsequently rotated with a half-wave plate to maximize the up-conversion efficiency. By comparing kinetics recorded parallel and perpendicular to the SF polarization axis, we resolve the degree of polarization (DOP) with sub-picosecond resolution and find it to remain constant throughout the emission burst. This observation indicates that linear polarization is established during the cooperative formation of the macroscopic SF dipole rather than developing after the onset of collective emission. Likewise, the fluence dependence of time-averaged emission spectra shown in Fig. 2E demonstrates that the degree of polarization (DOP) remains nearly constant from the SF threshold to excitation densities far above it, despite an almost two-order-of-magnitude increase in SF intensity. This is in stark contrast to previous work on chiral SF, which required high pump fluences to achieve circularly polarized emission (*11*). This robust performance shows that the cooperative build-up of progressively larger dipole ensembles strongly preserves the macroscopic linear polarization state.

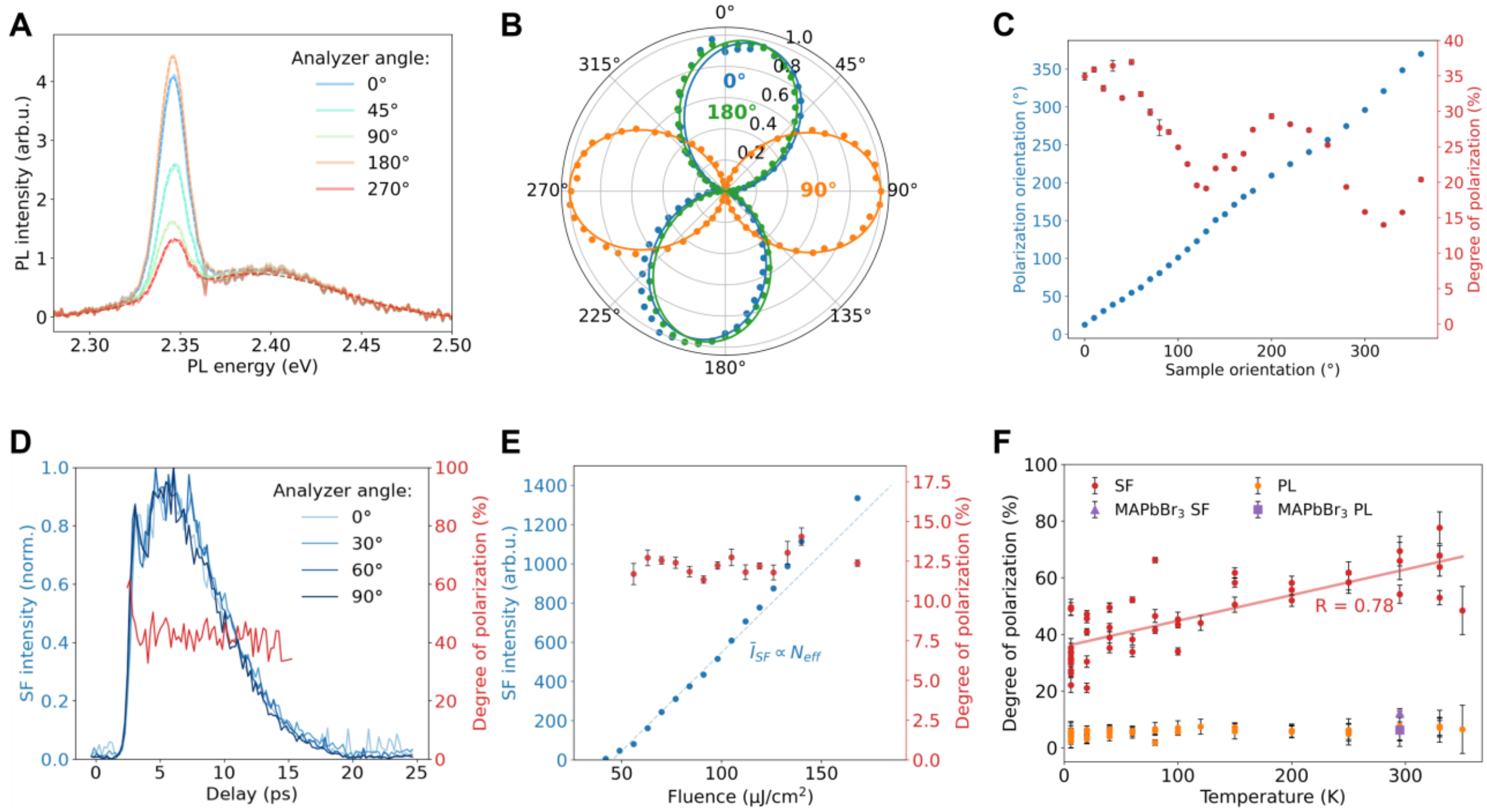

**Fig. 2. Polarization-resolved SF from *i*-PCPB thin films. (A)** Emission spectra recorded at different wire-grid analyzer angles. **(B)** Malus plots (normalized) of the SF intensity measured for different sample orientations at the same sample position. **(C)** Polarization angles of SF versus sample orientation. **(D)** Polarization-resolved SF sub-ps kinetics showing similar dynamics as function of analyzer angle and time-constant calculated DOP. **(E)** Fluence-dependent SF intensity and DOP. The dashed line indicates time-averaged SF scaling above threshold. **(F)** Temperature dependence of the DOP for SF and PL, including a three-dimensional mixed-cation lead-bromide perovskite for reference.

Thermal dephasing and carrier scattering are the primary reasons why SF is rarely observed in semiconductors at room temperature (*3*). Their influence on the polarization state of cooperative emission, however, remains largely unexplored. Surprisingly, our ferroelectric perovskite exhibits a systematic increase in the degree of polarization (DOP) with temperature (see Fig. 2F, Pearson correlation coefficient R = 0.78). The variance in the data arises from measurements at multiple excitation spots for each temperature (single spots at various temperatures shown in Section 13 in the supplement). The positive correlation between intrinsic polarization and lattice softening suggests that thermally activated dipolar dynamics interact with the cooperative build-up of SF, providing feedback that enhances the dominant polarization axis.

### Microscopic origin of linear polarization in superfluorescence process

As shown in Fig. 2, we observe linearly polarized SF in polycrystalline thin films, prepared by spin coating, featuring ferroelectric grain sizes of hundreds of nanometers (see Fig. S11 in the SI). Next, we study the impact of spatial heterogeneity on SF polarization to elucidate its underlying microscopic mechanism and to provide rationales for improving the DOP.

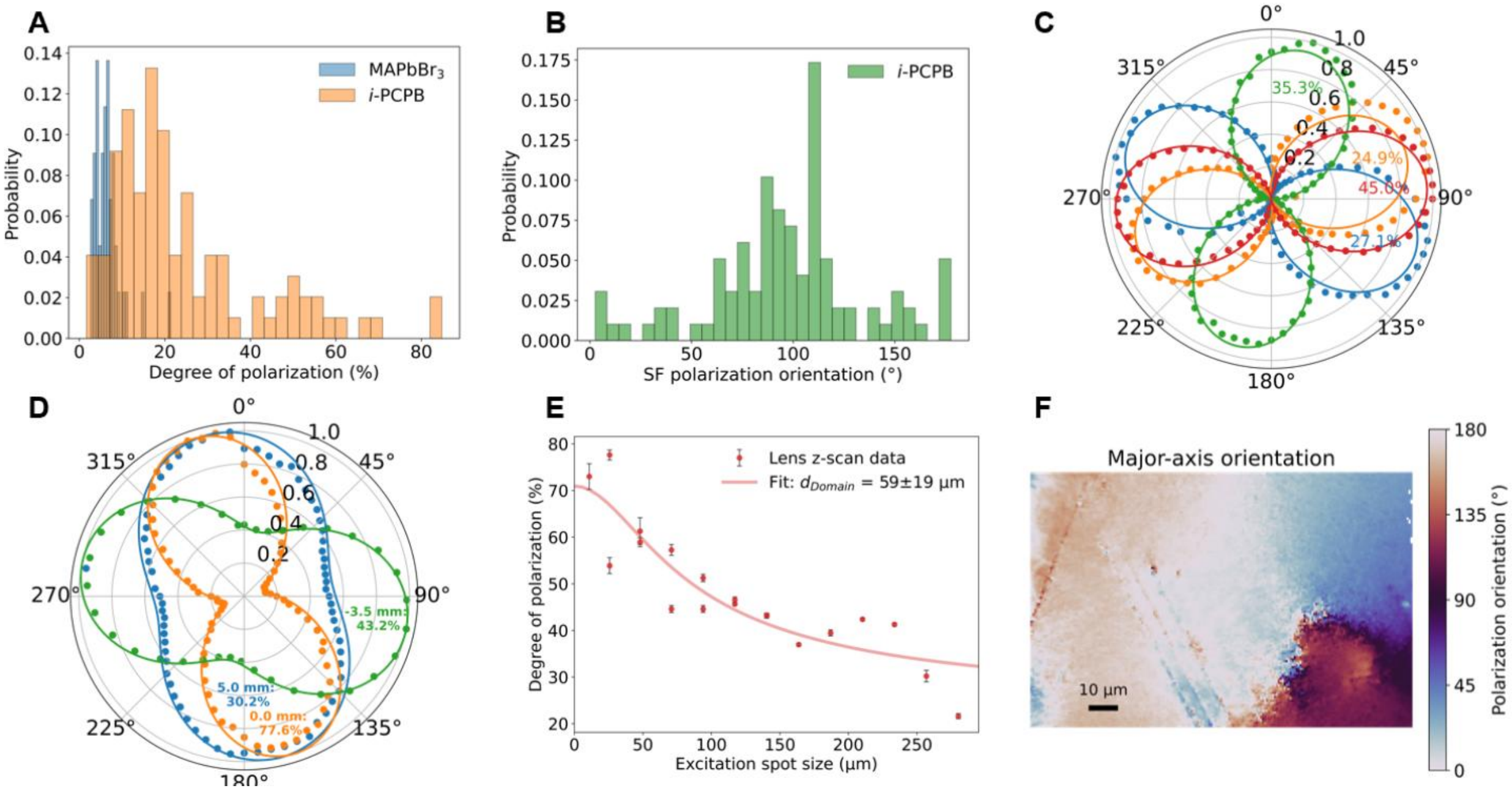

**Fig. 3 Local disorder effects in polarized superfluorescence. (A)** Histograms for DOP at several spots within an *i*-PCPB thin film (orange) and a 3D mixed-cation perovskite (blue). **(B)** Distribution of polarization orientation for *i*-PCPB regions shown in Figure A. **(C)** Polar plots (normalized to minimal and maximal SF intensity) with Malus fit for selected spots of the *i*-PCPB film from A and B. **(D)** Polar plots of SF intensity for varying excitation focus-sample distance. **(E)** Dependence of SF DOP and excitation spot size with a characteristic length fitted. Smaller excitation volumes at the pump beam focus correspond to larger DOPs. **(F)** Polarization sensitive microscopy of the SF emission shows multiple domains with distinct polarization orientations.

To ensure that linearly polarized SF is an intrinsic property of the *i*-PCPB thin film and not restricted to morphological outliers, we measured polarization resolved SF spectra for 100 randomly picked excitation spots (pump beam size ~200 μm). The resulting distribution of the DOP is shown in Fig. 3A in orange. The largest DOP we observe for a single spot is (86.5±1.9)% (see raw data and Malus fit in Fig. 10 in the SI). As a control, an identical experiment on a film of the 3D mixed-cation perovskite (MAGAFACs)$PbBr_3$, whose close analog $CsPbBr_3$ has recently been shown to exhibit SF (*13*), shows negligible DOP within our sensitivity levels. Because each spectrum represents an average over 400 excitation pulses, the negligible DOP observed for 3D perovskites could, in principle, arise from strongly polarized SF bursts with randomly fluctuating polarization axes. To test this possibility, we performed polarization-resolved single-shot spectroscopy. As shown in Section 9 of the Supplementary Materials, individual SF bursts from 3D perovskites are likewise unpolarized, demonstrating that the absence of macroscopic polarization is intrinsic rather than a consequence of averaging multiple bursts.

As expected for spin-coated polycrystalline hybrid perovskite films, different regions exhibit varying orientations of the linear polarization (shown in Figs. 3B and C) due to their respective local anisotropies. Despite this distribution being not entirely random (Fig. 3B), reflecting a weak preferential in-plane alignment of grains and ferroelectric domains during film formation, regions deviating from the more often observed major polarization axis show similar DOPs, ruling out systematic biases in the optical setup (see Fig. 3C).

The unusually high DOP of SF compared with both PL and ground-state absorption suggests that the cooperative build-up of SF amplifies weak local anisotropies within the excitation volume. If so, the DOP should increase for smaller excitation spots, in which the number of anisotropic domains or grains that are spatially averaged is decreased, and polarization-seeding anisotropies should be increased. To test this hypothesis, we varied the excitation spot size while maintaining the collection geometry. Figures 3D and 3E show a pronounced increase in DOP for smaller excitation areas, consistent with reduced averaging over independently oriented domains. However, the characteristic length of the DOP with respect to the excitation spot size (see fit line in Fig. 3E) of 60 µm is far larger than the elementary unit of anisotropy, that is single grains of the polycrystalline film. To further elucidate the origin of ensemble averaging, we conducted polarization-resolved wide-field microscopy of PL and SF (see details in Section 15 in the SI). The polarization orientation was calculated for each pixel and is displayed in Fig. 3F. The presence of several regions (~ 50-100 µm in size) with uniform yet distinct polarization orientation suggests that multiple SF bursts with individual polarization state, determined by their local anisotropy, emerge independently in spatially and thus temporally separated regions and cause the ensemble averaging observed in Figs. 3D and 3E. This finding suggests that larger regions of common anisotropy, realized through growth of larger grains or ferroelectric domains provide a direct route toward brighter, highly linearly polarized SF.

To place our observations within the established theoretical framework of superfluorescence developed by Bonifacio et al (*29*), we extend the semiclassical Maxwell–Bloch model by allowing the coupling of each transition dipole to the common radiation mode to depend on its orientation, as expected for anisotropic excitons in a two-dimensional semiconductor. The model reproduces the characteristic scaling of the SF burst with excitation density and demonstrates that weak initial anisotropies in the dipole-orientation distribution, corresponding to only a few percent DOP in spontaneous emission, are sufficient to generate strongly linearly polarized SF bursts with average degrees of polarization comparable to those observed experimentally (20–90%; Fig. S21 and Section 18 in the SI). These simulations suggest that the coherent build-up of the macroscopic polarization can amplify subtle structural anisotropies into robust macroscopic polarization properties of the emitted light. Notably, the same model also predicts linearly polarized individual SF bursts for the three-dimensional lead-bromide perovskite, in contrast to our single-shot measurements (see Fig. S8). This discrepancy indicates that orientation-dependent dipole coupling is an unsuitable concept for three-dimensional lead halide perovskites and highlights the distinct role of reduced dimensionality and excitonic anisotropy in the ferroelectric two-dimensional material.

## Conclusion

Our results show that superfluorescence can act as a cooperative amplifier of weak material anisotropies. In a polycrystalline ferroelectric layered perovskite, we observe strongly linearly polarized SF with DOPs up to 86%, despite negligible linear dichroism and weakly polarized spontaneous emission. The polarization emerges early in the cooperative build-up of the SF state and is robust to spatial averaging over locally anisotropic regions. Together with our modified Maxwell–Bloch model, these findings show that small microscopic biases can be amplified into robust macroscopic optical order, enabling strong polarization performance even when long-range microscopic material alignment and structural control remain incomplete. More broadly, this establishes SF as a sensitive probe of material anisotropies that remain nearly invisible to linear spectroscopy or spontaneous emission, linking microscopic symmetry breaking to emergent macroscopic optical order.

**Acknowledgments:**

**AI use statement:** During manuscript preparation, ChatGPT-5.5 (OpenAI) and Claude Sonnet 5 were used to assist with coding, language editing and the restructuring of text for clarity and readability. Likewise, the authors reviewed, revised, and verified all generated suggestions. All scientific ideas, analyses, interpretations, and conclusions originate from the authors, who take full responsibility for the manuscript.

The authors thank Tiago Buckup for technical support.

**Funding:**

F.D. acknowledges funding by the German Research Foundation (DFG) via Research Training Group GRK 2948/1.

F.D. acknowledges funding from the Athenaeum—Dietrich Götze Stiftung für Kultur und Wissenschaft and start-up funding from Heidelberg University.

C.G. acknowledges financial support from the China Scholarship Council (CSC).

Part of this research has been funded by the Deutsche Forschungsgemeinschaft (DFG, German Research Foundation) – project number 545050087.

**Author contributions:**

Conceptualization: FD, CG

Methodology: CG, DS, FD, JT, SK

Investigation: CG, DS, TRE, JT, SK

Validation: TRE, DS

Visualization: DS, CG, TRE

Funding acquisition: FD

Project administration: FD

Supervision: CG, DS, FD

Writing – original draft: DS, FD, CG

Writing – review & editing: DS, FD, TRE

**Competing interests:** Authors declare that they have no competing interests.

**Supplementary Information**

Materials and Methods

Supplementary Text

Figs. S1 to S22

Supplementary Information for

# Superfluorescence as a cooperative amplifier of hidden anisotropy in a ferroelectric hybrid perovskite

Changhao Gao[1]†, Daniel Sandner[1]†*, Talia Ruhrberg Estévez[1], Jeron Timmer[1], Sophia Klubertz[2], Martijn Kemerink[2], Felix Deschler[1*]

[1] Physikalisch-Chemisches Institut, Universität Heidelberg, 69120 Heidelberg, Germany

[2] Institute for Molecular Systems Engineering and Advanced Materials, Universität Heidelberg, 69120 Heidelberg, Germany

† These authors contributed equally to this work

*Corresponding authors, Email: daniel.sandner@uni-heidelberg.de, deschler@uni-heidelberg.de

**The PDF file includes:**

## Section 1: Sample synthesis and preparation

### Single Crystal Preparation

*i*-$PA_2PbBr_4$ single crystals were synthesized by a controlled-cooling method. PbO (1.00 g) and isopentylamine were added in a stoichiometric ratio of 1:2 to 15 mL of hydrobromic acid (48 wt%). The mixture was heated until completely dissolved and maintained at 85 °C for 24 h. The solution was then cooled to room temperature at a rate of 1 °C $h^{-1}$. The resulting crystals were washed three times with toluene and dried under vacuum at room temperature for 24 h before being transferred into a glovebox.

### Preparation of *i*-PCPB thin films

Glass substrates were sequentially ultrasonicated in deionized water, acetone, and isopropanol for 10 min each, followed by UV-ozone treatment for 30 min. The precursor solution was prepared by dissolving *i*-$PA_2PbBr_4$, CsBr, and $PbBr_2$ in anhydrous dimethyl sulfoxide (DMSO) at a molar ratio of 0.2:1.1:0.8. The final solution volume was adjusted to give a total Pb concentration of 0.30 M. The solution was stirred at 60 °C for 2 h in an N2 glovebox. For film deposition, 80 µL of the precursor solution was dispensed onto each glass substrate and spin-coated at 3000 rpm for 2 min. The resulting films were annealed at 80 °C for 10 min. Precursor preparation, spin coating, and annealing were performed in a glovebox.

### Preparation of mixed-cation reference films

A mixed-cation bromide precursor was obtained by combining methylammonium bromide (MABr, 27.2 mg), guanidinium bromide (GABr, 67.2 mg), cesium bromide (CsBr, 78.8 mg), formamidinium bromide (FABr, 212.0 mg), and lead bromide ($PbBr_2$, 881.0 mg) in 3 mL of dimethyl sulfoxide. The mixture was heated at 80 °C until a homogeneous solution was obtained. Films were deposited on glass substrates by spin coating at 4000 rpm for 35 s. With 10 s remaining in the spinning cycle, 150 µL of chlorobenzene was applied to the rotating substrate as an antisolvent. The as-deposited films were neither thermally annealed nor encapsulated.

### PMMA encapsulation

A 1.5 wt% solution of PMMA in anhydrous chlorobenzene was stirred at 60 °C until completely dissolved. After annealing, 400 µL of the PMMA solution was deposited directly onto each *i*-PCPB film and spin-coated at 1000 rpm for 1 min. The encapsulated films were subsequently annealed at 80 °C for 10 min. PMMA encapsulation was performed in a glovebox. All samples used for optical measurements were encapsulated with PMMA, whereas those used for piezoresponse force microscopy were left unencapsulated.

## Section 2: Spectroscopy of SF

### Description of up-conversion setup:

We measured time-resolved photoluminescence and superfluorescence kinetics using a commercial HALCYONE Fluorescence up-conversion Spectrometer (by Ultrafast systems, Sarasota, FL, 34240 USA) powered by an amplified Ti:Sa laser system (Coherent Astrella, 800 nm, 4 kHz, 90 fs). Briefly, the sample is excited by a 400 nm beam and emitted light is collected by an off-axis parabolic mirror. Emitted light and an 800 nm gate beam are spatially overlapped in a BBO crystal. The upconverted light in the uv range is recorded by a spectrometer (pixel array) as a function of the BBO crystal angle and time-delay of the gate pulse.

Note that the up-conversion process is intrinsically polarization sensitive through the $\chi^2$ tensor and phase matching in the birefringent crystal. Thus, intentional polarization resolution requires analyzers and halfwave plates (HWP). We obtained polarization-dependent kinetics by filtering the polarization state of the emitted light and then optimized the up-conversion efficiency using the HWP. The angle of the HWP matched the expectations.

**Description of time-averaged SF spectroscopy**

Time-averaged PL spectra were recorded with an Avantes Starline spectrometer through a multimode optical fiber cable. For all experiments except temperature dependence and single-shot experiments, the aforementioned Coherent Astrella was used for pumping with the second harmonic. For temperature-dependent experiments and the single-shot polarization analysis of the 3D perovskite, we excited the samples using also at 400 nm with the output of an OPA (Orpheus, light conversion) powered by an ytterbium-based ultrafast solid-state regenerative amplifier (Pharos, light conversion) operating at variable repetition rate. Temperature dependence was measured by mounting the sample onto a closed-cycle He cryostat (Montana Instruments).

**Section 3: Piezoresponse force microscopy and hysteresis curves:**

Piezoresponse force microscopy measurements were performed using an Asylum Research Jupiter atomic force microscope operated in contact-mode dual AC resonance tracking PFM (DART-PFM). A nominal contact-force setpoint of 5 nN was used. For spatially resolved imaging, a 5.5 × 5.5 $\mu m^2$ area was scanned at 256 × 256 pixels with a scan rate of 1.91 Hz. Two AC excitation frequencies of approximately 377.47 and 387.47 kHz, separated by 10 kHz, were applied with a drive-amplitude setting of 4.02 V for each excitation channel. The surface topography, two amplitude signals, two phase signals, and the tracked contact-resonance frequency were recorded simultaneously.

Local piezoresponse hysteresis measurements were subsequently performed at selected positions using DART-PFM switching spectroscopy. The DC bias was cycled between approximately −7 and +7 V for three consecutive cycles using a triangular switching waveform at 0.243 Hz. The DART amplitude, phase, and contact-resonance frequency were recorded as functions of the applied bias at a sampling rate of 2000 points $s^{-1}$ and a detection bandwidth of 1 kHz. The hysteresis analysis was based on the off-field response measured after removal of each DC-bias pulse, thereby reducing electrostatic contributions to the measured piezoresponse.

**Section 4: Basic characterization**

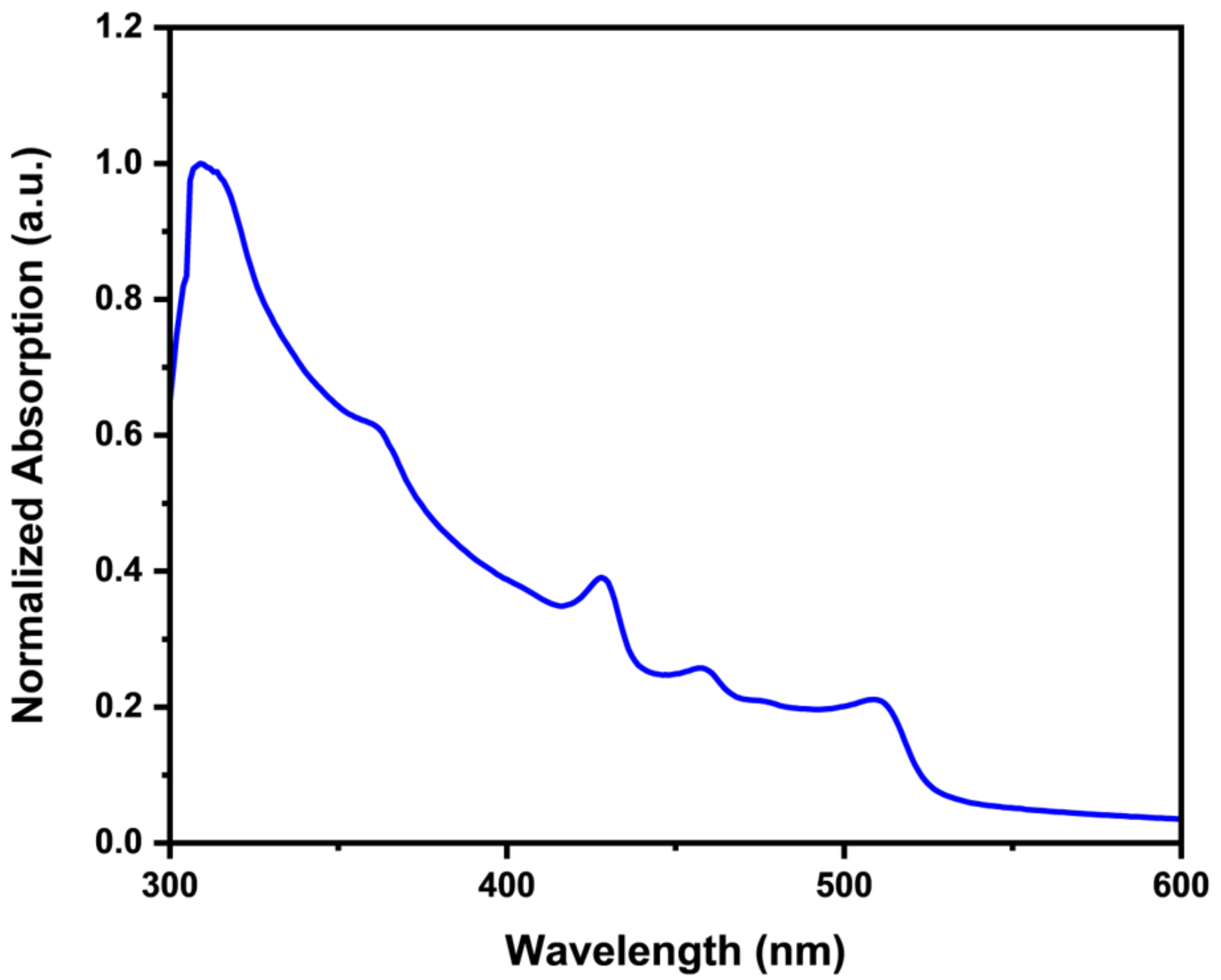

Fig. S1: The UV-Vis absorbance spectrum of *i*-PCPB.

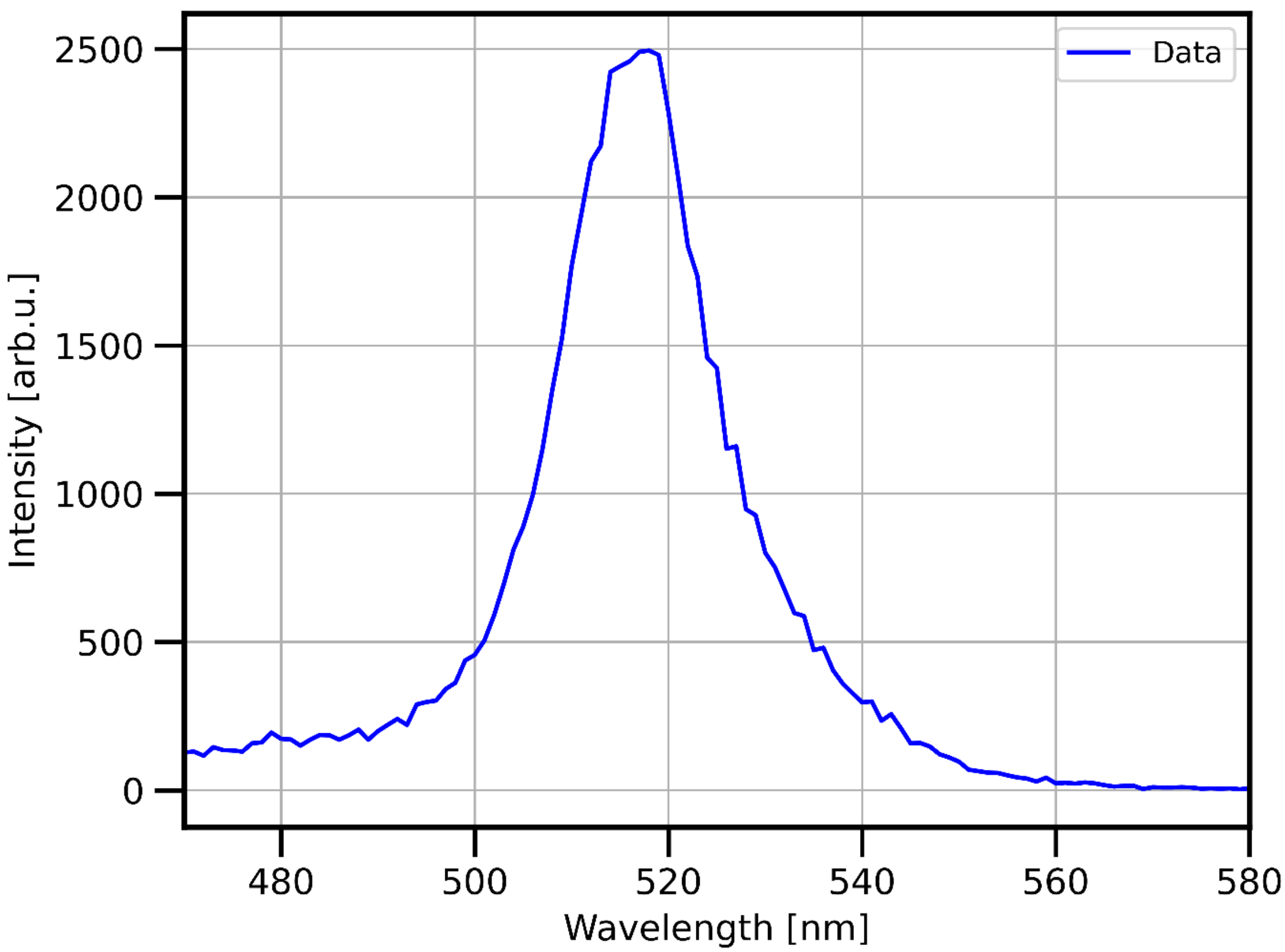

Fig. S2: PL spectrum of i-PCPB under steady-state excitation.

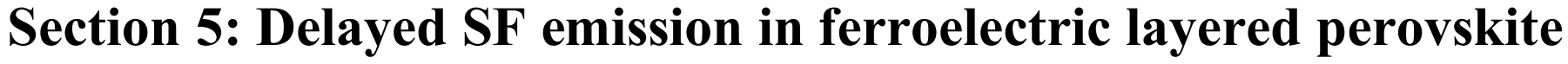

**Section 5: Delayed SF emission in ferroelectric layered perovskite**

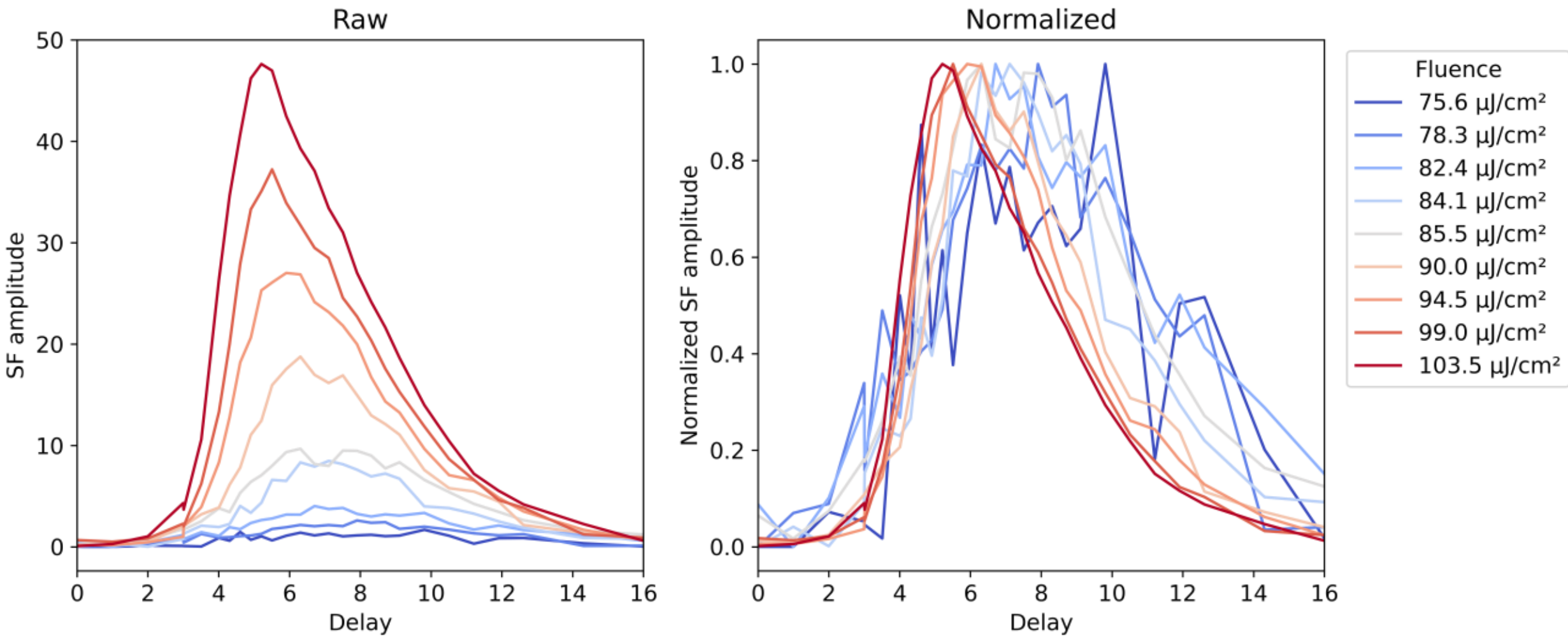

Fig. S3: SF up-conversion kinetics (unnormalized left, normalized, right) for *i*-PCPB thin films excited with various fluences at 400 nm.

One can clearly see the super-linear increase in SF peak intensity and the increase in emission-delay with decreasing fluence, characteristic for the Dicke model of superfluorescence.

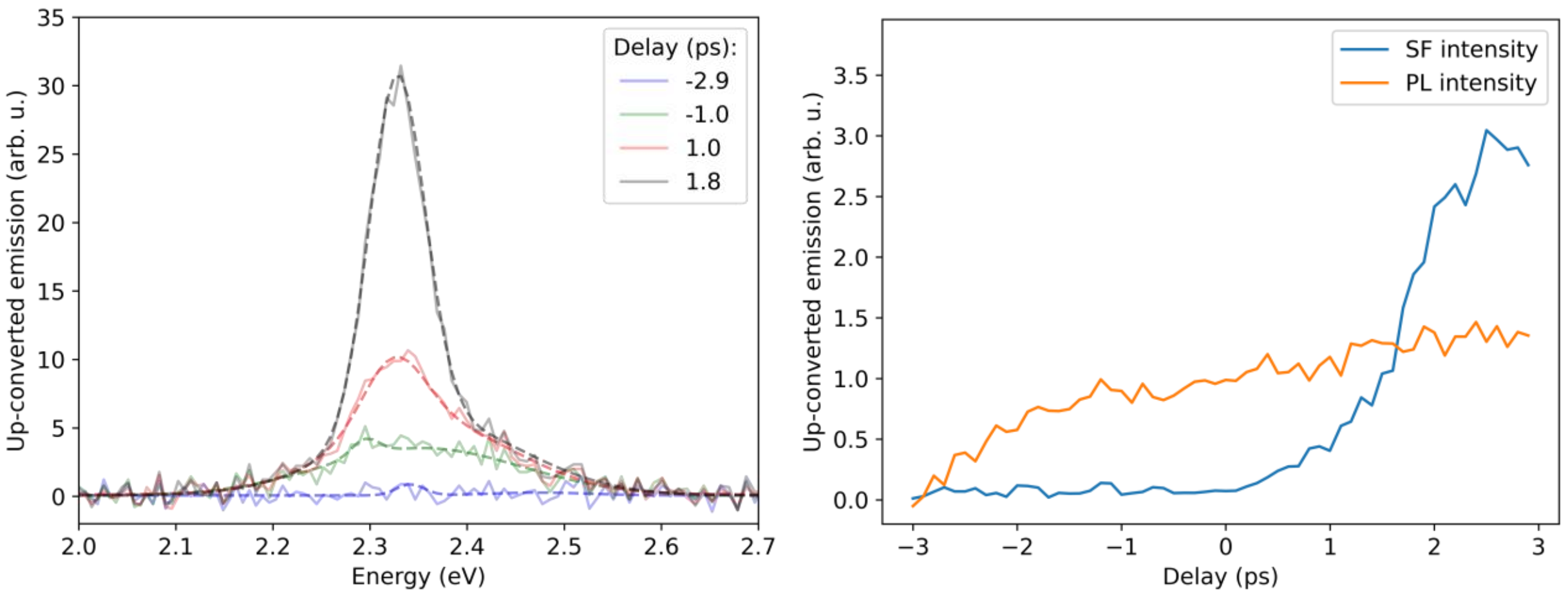

Fig. S4: Decomposition of up-converted emission spectra (left) into broad PL and narrow SF, each represented as gaussian peaks (dashed lines are fits). Right side: Kinetics of the fitted amplitudes of SF and PL, clearly showing delayed emission.

Spontaneously emitted PL and SF can be distinguished by their spectral width. Applying this fitting procedure to up-converted emission spectra with few 100 fs time-resolution clearly shows the delay between PL and the SF burst (Fig. S4 right side). The rise of PL after excitation can be explained by hot carrier cooling and charge carrier transfer between 2D and 3D phases.

**Section 6: Characterization of absorption anisotropy:**

To test the optical polarization anisotropy close to the bandgap in regions like our pump-spot size, we performed micro-uvvis spectroscopy with a homebuilt setup. We investigated the same region as used for the orientation dependence in Fig. 2. The sample was mounted directly onto a 100 µm spherical pinhole (Thorlabs) such that all excitation light and the probe beam of the uv-vis spectrometer probe the same region.

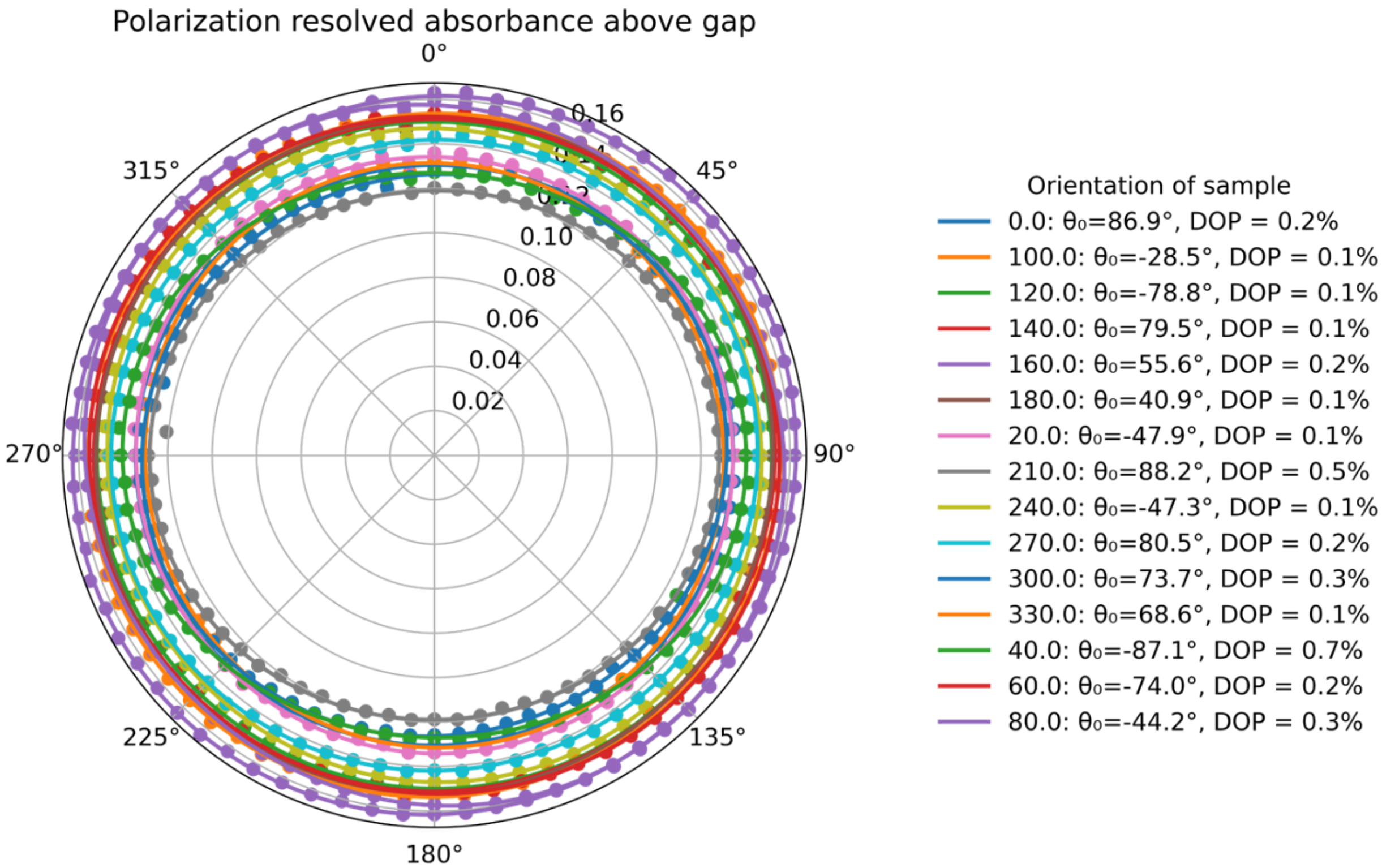

Fig. S5: Polarization resolved above-gap absorption of an *i*-PCPB thin film for the same region (100 µm diameter) as presented in Figs. 2 b and c.

Contrary to SF, ground state absorption at the 3D phase exciton resonance of 2.4 eV (~520 nm, see Fig. S1), which is the electronic state of the SF emission we observe, shows negligible degrees of polarization and a lack of relation between sample orientation and absorption anisotropy. Together with the weak PL polarization, we can conclude that the studied region comprises many randomly oriented grains which are spatially washed out in conventional spectroscopy methods. We demonstrate below in Section 18 that the spontaneously synchronized macroscopic dipole in SF is susceptible to the remaining small imbalances and amplifies it into a macroscopic linearly polarized light emitting state.

**Section 7: Polarization resolved sub-ps SF kinetics**

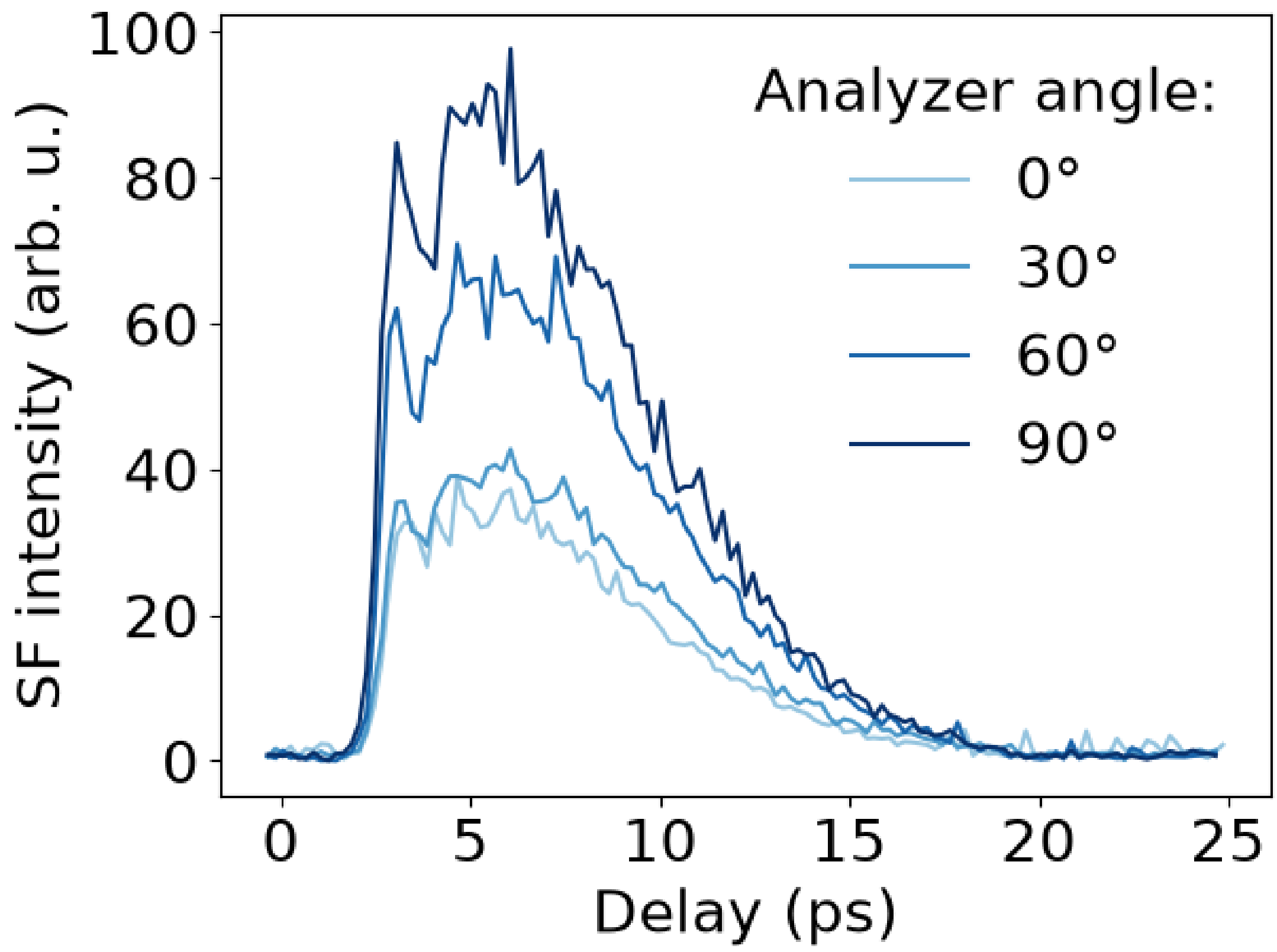

Fig. S6: Unnormalized polarization-resolved SF kinetics for different analyzer angles. Basis to the ultrafast DOP data presented in Fig. 2D.

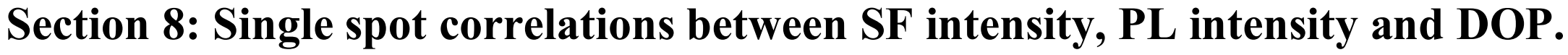

**Section 8: Single spot correlations between SF intensity, PL intensity and DOP.**

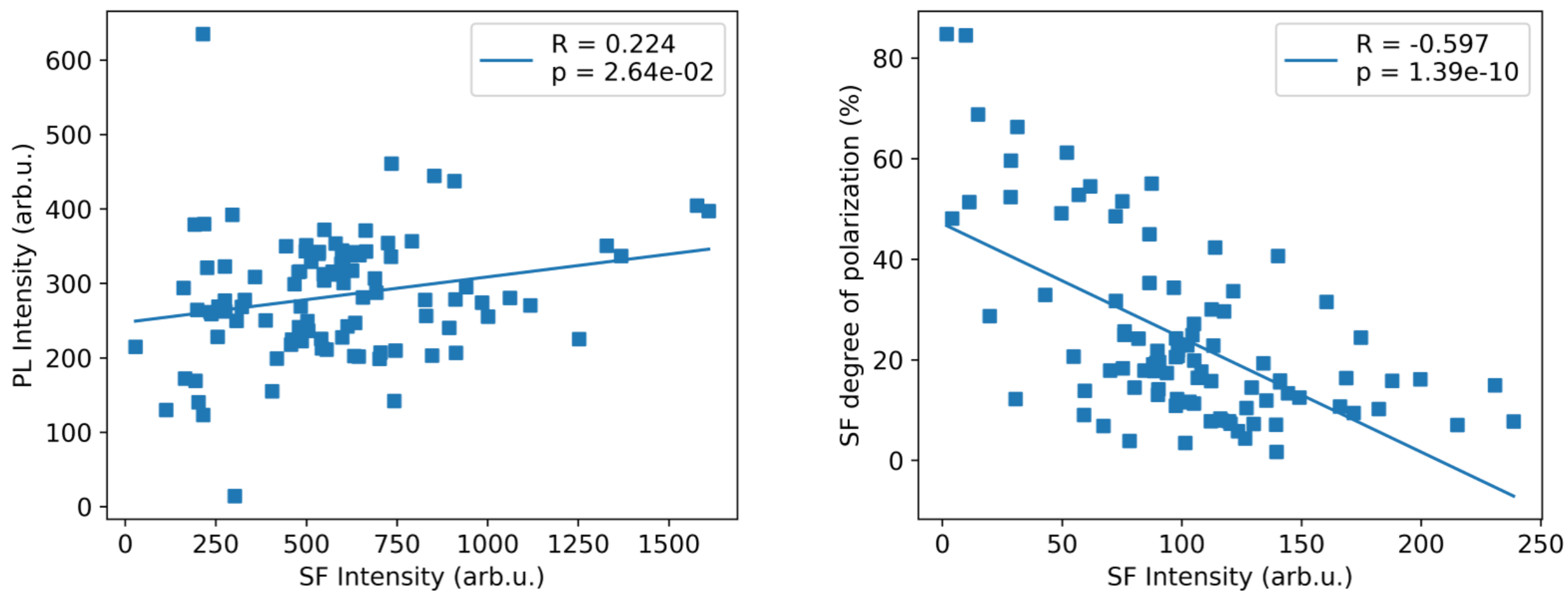

Fig. S7 Correlations between various properties obtained by fitting the polarization-resolved emission spectra above SF threshold for many spots (histograms of DOP shown in Fig. 3A)

We only find weak correlations between the PL intensity and SF intensity for different spots of the *i*-PCPB thin film. This is partially expected, since the PL yield depends on phenomena on the ns-µs timescale governed by trap densities. Instead, superfluorescence bursts occur fully within the first 20 ps and are thus less influenced by diffusion or trap states, yet the required synchronization/rephasing requires the suppression of carrier-carrier and carrier-phonon scattering. The remaining weak correlation can be explained by variations in the film thickness or the absorption coefficient at the pump energy (3.1 eV).

However, there is a weak, yet negative correlation between the SF DOP and SF intensity. We can only speculate about the origin of this correlation but the data presented in Fig. 3e-f support the following hypothesis: In case only a single domain exceeds the excitation threshold and participates in SF, then weaker SF intensity is observed albeit with larger DOP due to the lack of ensemble averaging of multiple superfluorescent domains with distinct polarization states.

**Section 9: Single-shot resolved SF in 3D cubic perovskites**

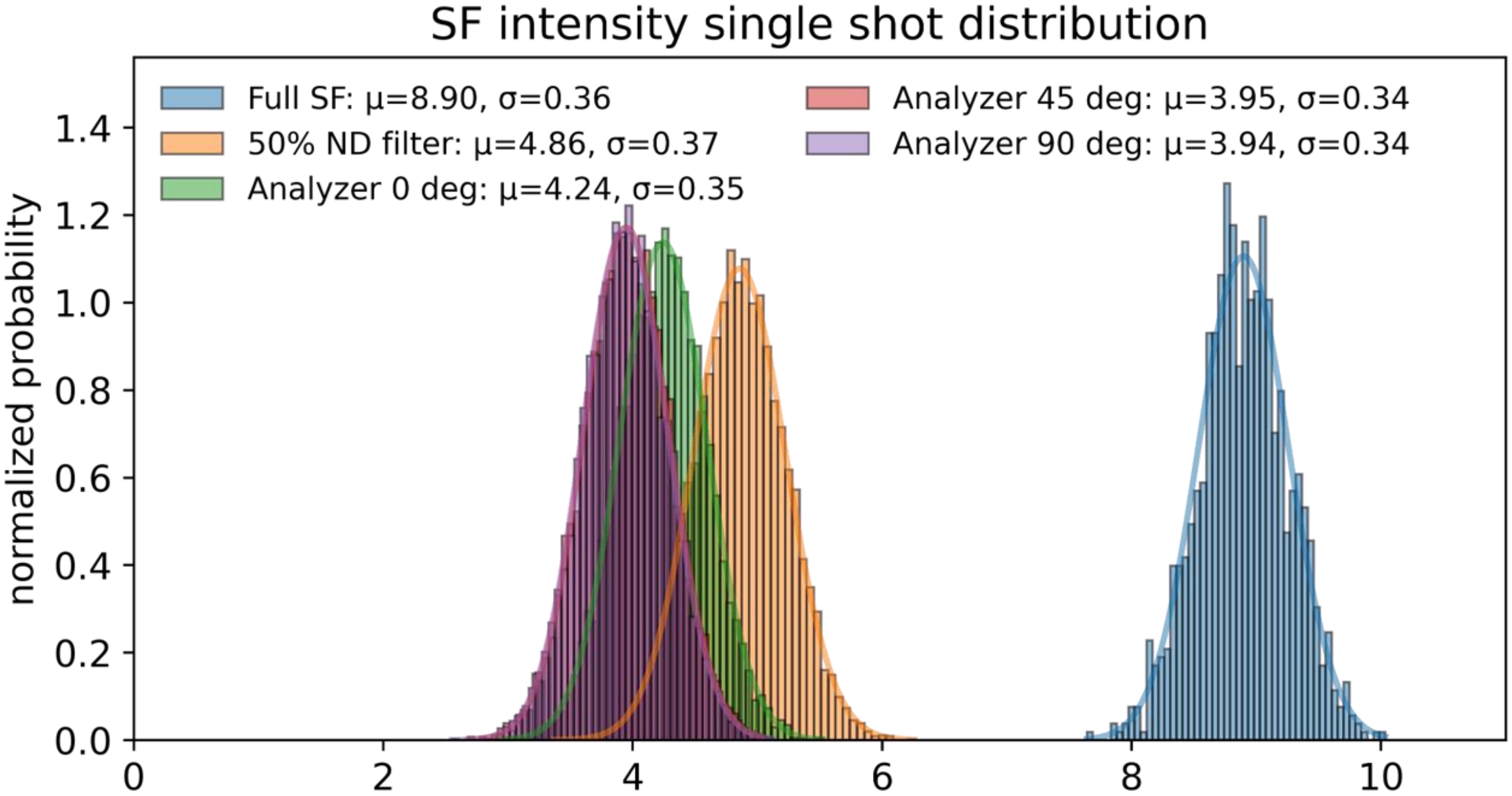

Fig. S8: Intensity distributions of each 10.000 SF bursts recorded for $MAPbBr_3$ without polarizing filters, with wire-grid polarizers and 50% transmitting polarization-insensitive ND filter. For linearly polarized SF bursts with random orientation, the intensity distribution after a linear polarizer would expand from zero to the intensities obtained without polarizer (with the effect diminished by smaller DOP or unpolarized components). The results and statistics presented in Figure S8 indicate that even single SF bursts of the 3D perovskite show less than 5% DOP. Thus, strongly linearly polarized SF is observed in anisotropic i-PCPB but not in the three-dimensional control under otherwise comparable conditions.

The consequences are discussed in detail in the modeling section of polarized SF below.

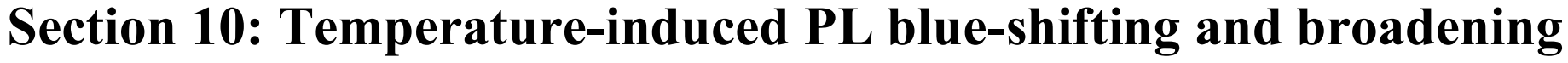

**Section 10: Temperature-induced PL blue-shifting and broadening**

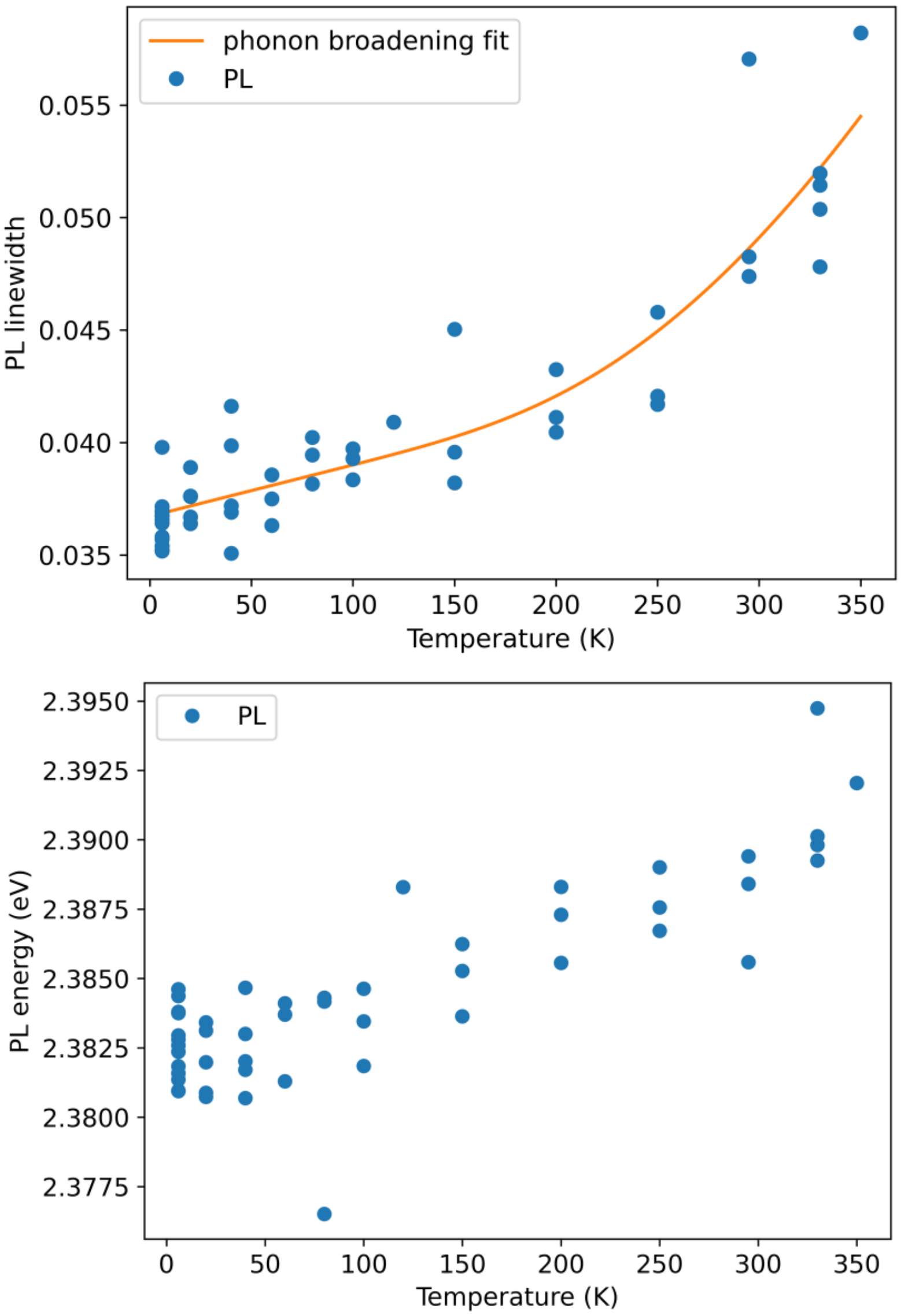

Fig. S9. Width (top) and center energy (bottom) of PL versus temperature.

As commonly observed for semiconductors, we find the width of PL to increase with temperature. The orange line is a fit according to the common model:

$$\Gamma(T) = \Gamma_0 + \frac{\Gamma_{LO}}{\exp\left(\frac{E_{LO}}{k_B T}\right) - 1}$$

However, the uncertainties are too large to identify the dominant contribution in phonon broadening.

Likewise, we confirm the inverse Varshni shift; an increase in bandgap with temperature; which is observed for most lead-halide perovskites. The absence of abrupt jumps in the emission energy indicates a single crystallographic phase in the studied temperature range.

**Section 11: Largest reported DOP**

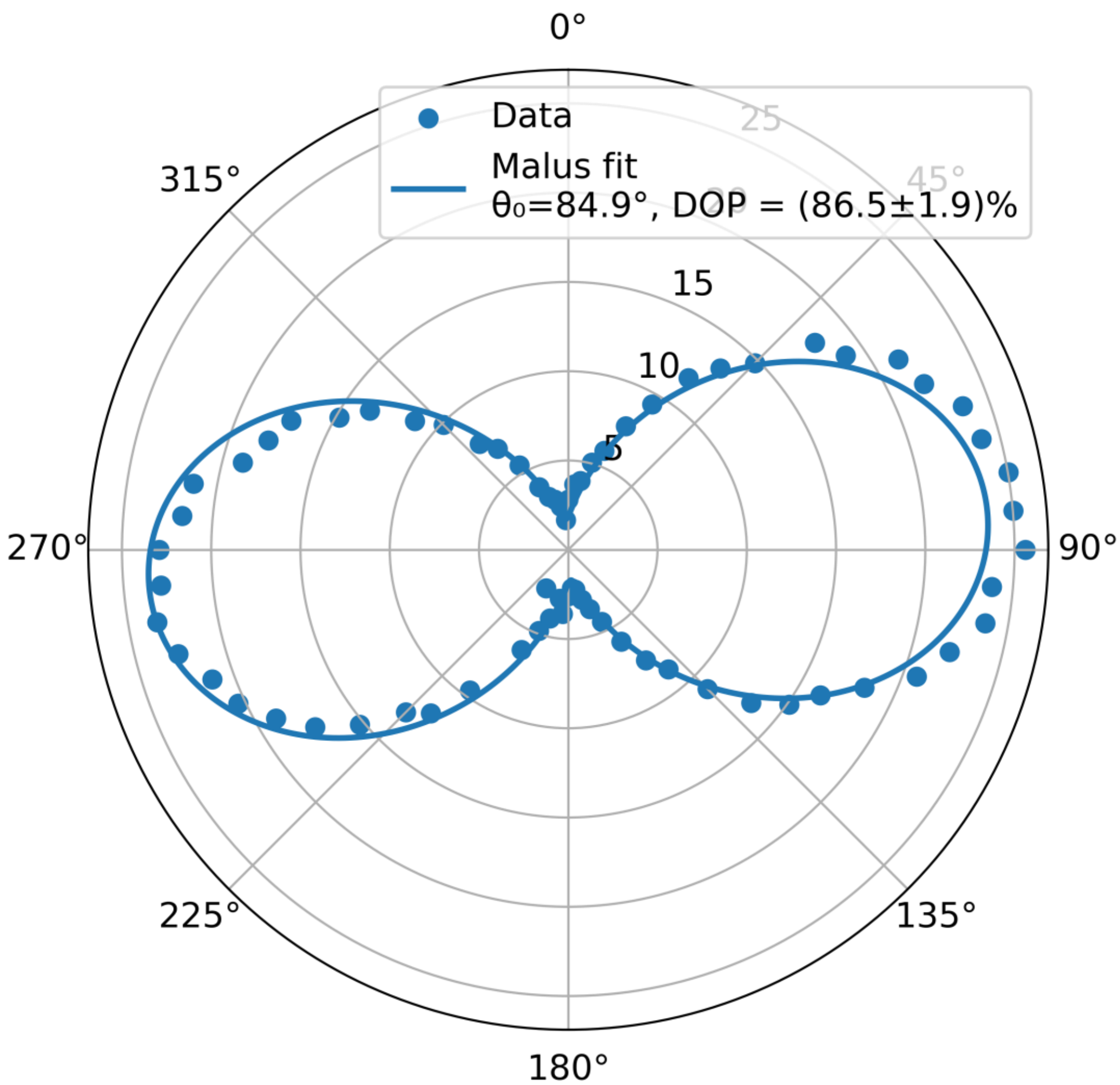

Fig. S10: Record holder of spot-statistics shown in Fig. 3A. The unnormalized polar plot of SF intensity with Malus fit supports the reported DOP of 86%.

## Section 12: PFM imaging and switching

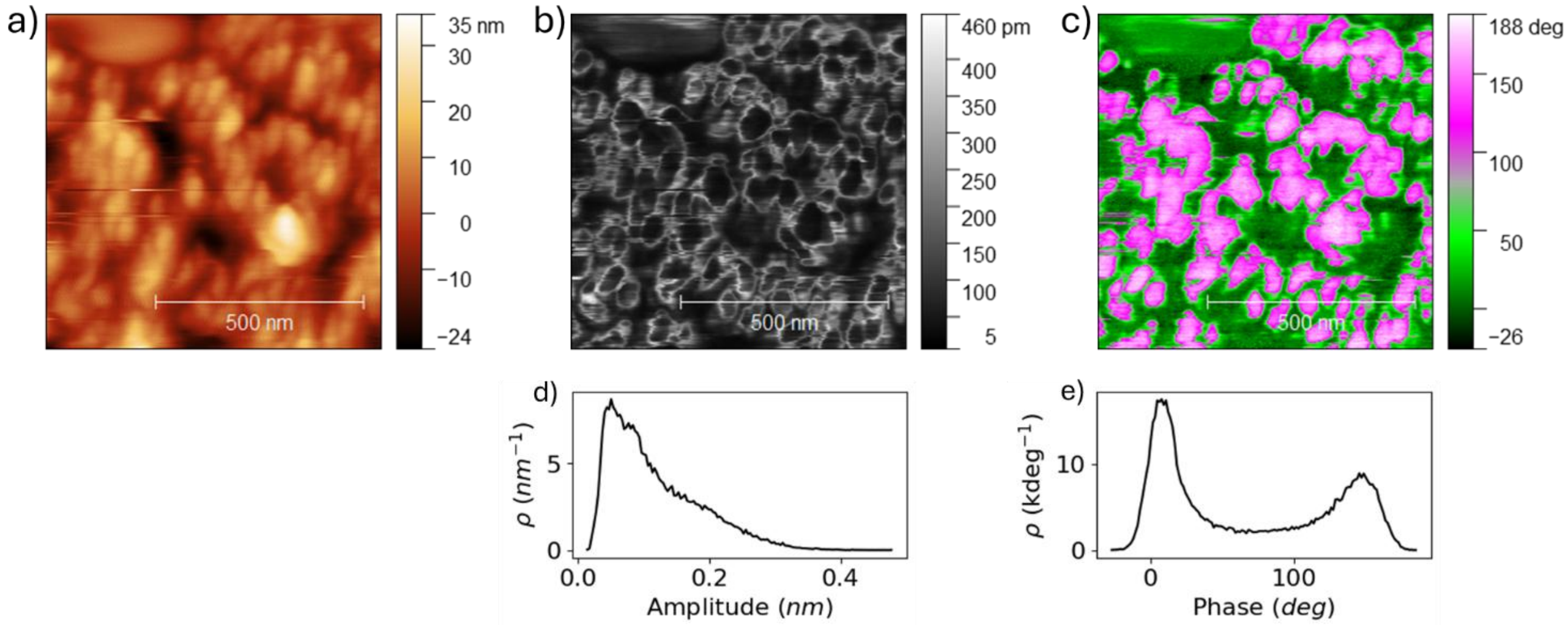

Figure S11: Dual-AC resonance tracking (DART) PFM measurements on an i-PCPB thin film, spin-coated on an ITO coated glass substrate. Measurements were carried out at room temperature under ambient conditions: a) topography, b) amplitude, c) phase signal. Distributions of the amplitude and phase signal are depicted in d) and e).

Piezo Response Force Microscopy (PFM) is a powerful technique for probing the local piezoelectric response of a material, thereby revealing the presence of ferroelectric domains with different orientations of spontaneous polarization within the film. Since the probing field between AFM-tip and substrate is cross-plane, we show the component of the piezoresponse orthogonal to the substrate.

The amplitude and phase images indeed reveal distinct regions corresponding to domains with different piezoresponses. The phase difference between these regions is close to 180° (see the two distinct peaks in panel (e)), as expected for ferroelectric domains with polarization oriented either towards or away from the sample surface. The amplitude distribution (panel d)) is dominated by low values, indicating that most domains are preferentially oriented in-plane and therefore exhibit only a small out-of-plane piezoresponse.

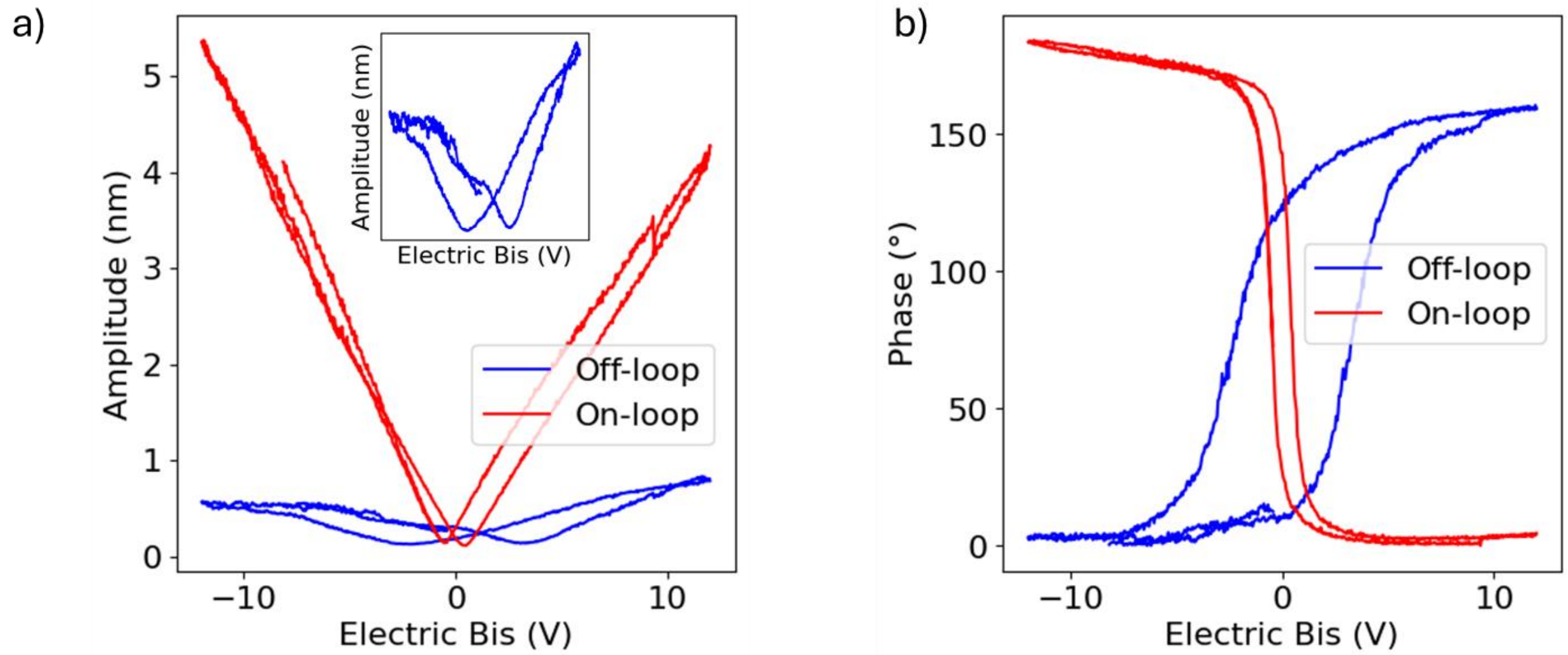

Figure S12: PFM-spectroscopy on- and off-loop measurements of the i-PCPB thin film with a) amplitude, inset shows the off-loop measurement magnified for better visualization, and b) phase measurement.

For the PFM spectroscopy, an AC bias sweep with a maximum amplitude of 12 V was applied using alternating on- and off-field states. While the on-field response is dominated by electrostatic interactions between the tip and the sample, the off-field measurements exhibit the characteristic butterfly-shaped amplitude loop and a saturated phase–bias hysteresis, both of which are hallmarks of ferroelectric switching. The approximately 180° phase contrast between oppositely poled states further confirms reversible polarization switching, with coercive voltages of approximately 3 V.
In combination with the results presented in Fig. S11, these measurements demonstrate the ferroelectric and piezoelectric nature of i-PCPB.

## Section 13: Temperature dependence of the degree of polarization

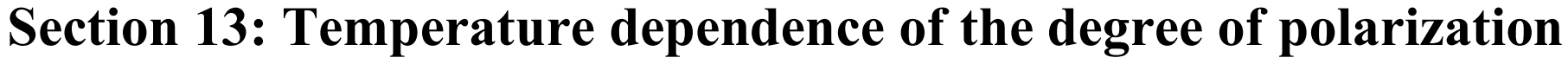

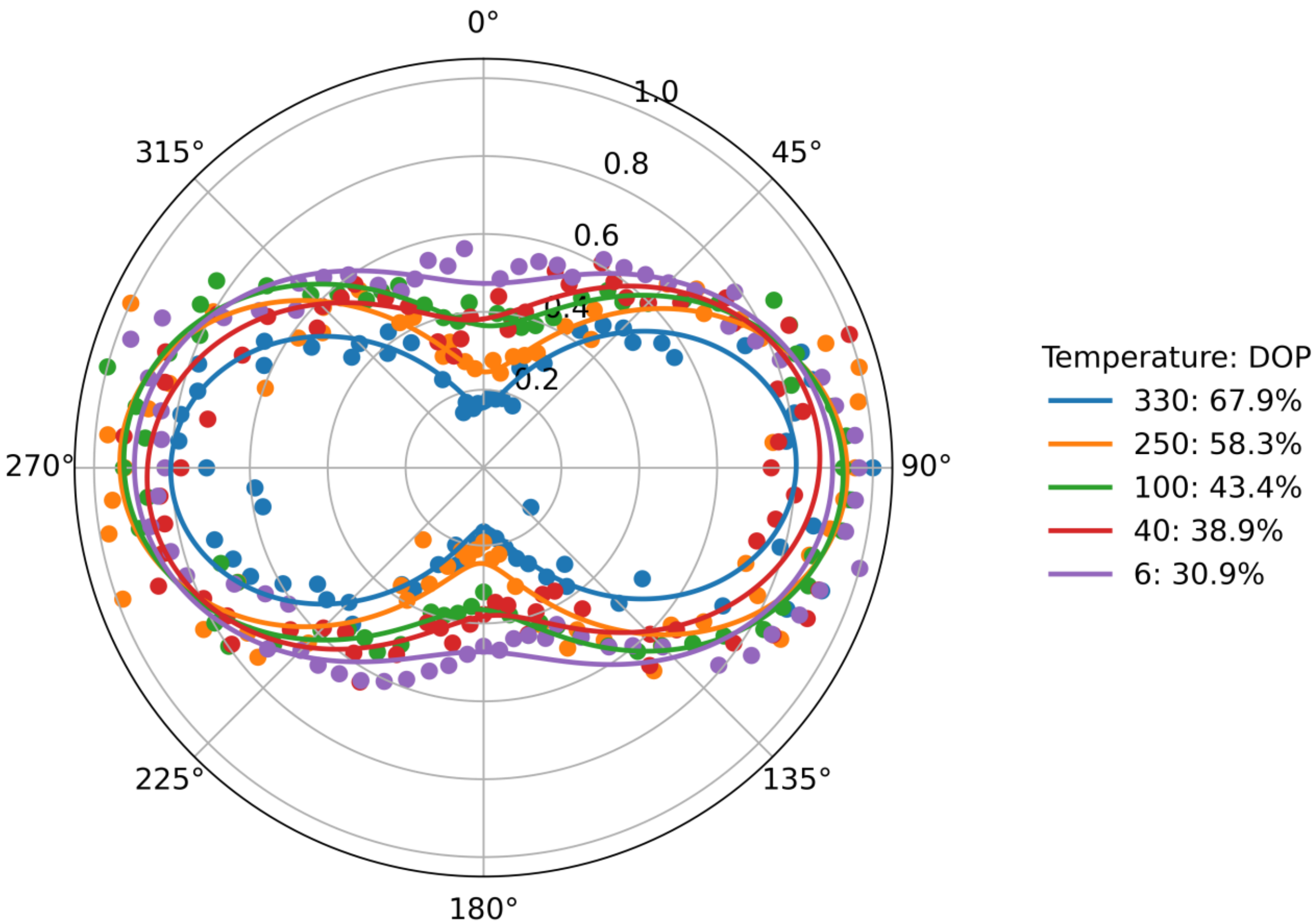

Fig. S13: Polarization resolved SF of a single region (~100 µm pump diameter) at various temperatures

The temperature dependence presented in Fig. 2F shows different regions and at different temperatures. Fig. S13 confirms the trend of increasing DOP with temperature for a single 100 µm sized region, avoiding the heterogeneity involved in comparing different regions.

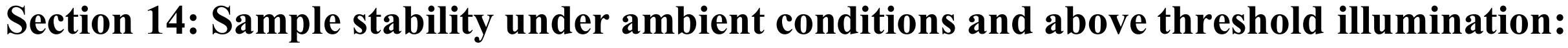

**Section 14: Sample stability under ambient conditions and above threshold illumination:**

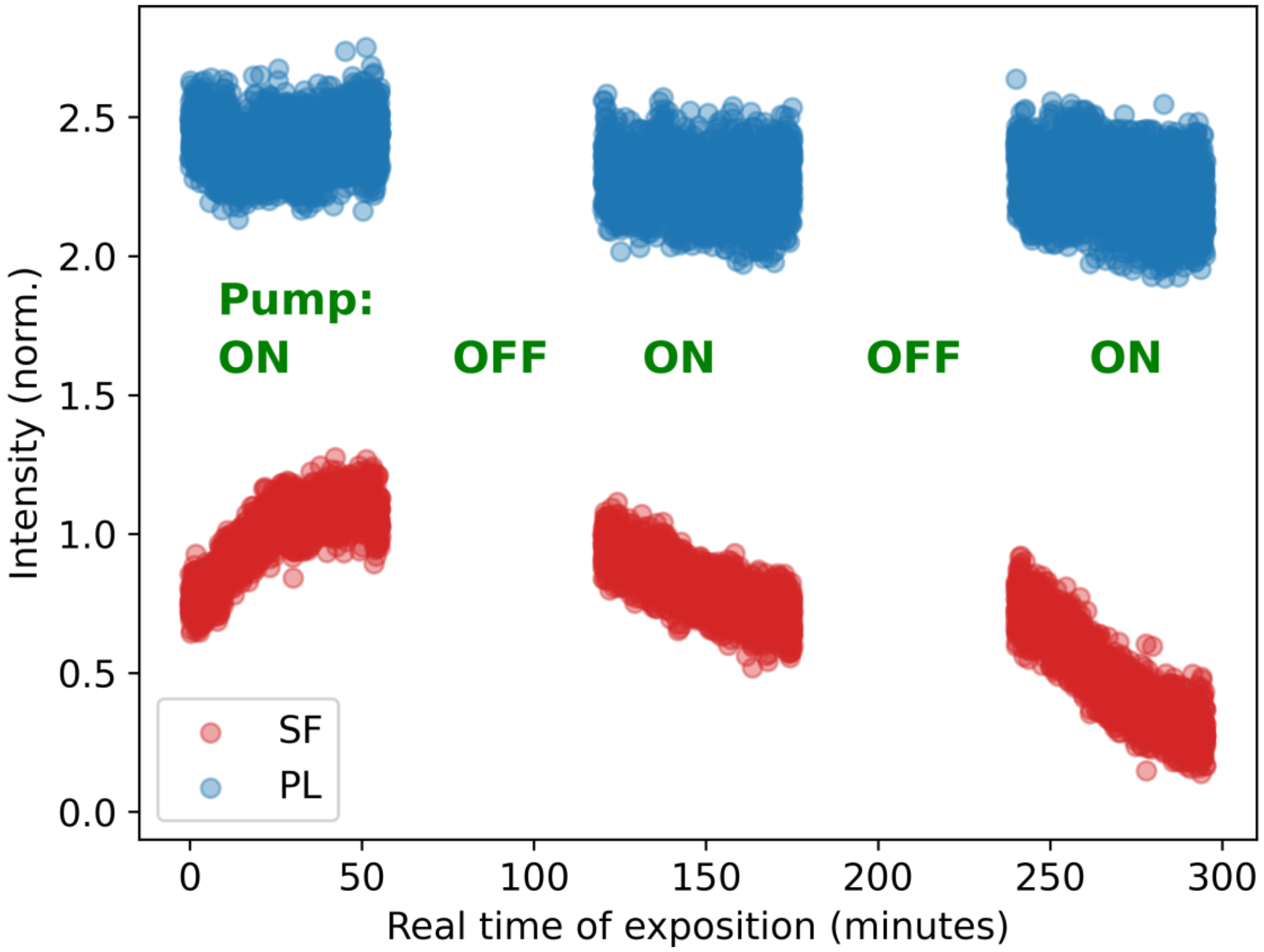

Fig. S14: SF intensity versus real-time of exposure at a laser-repetition rate of 4 kHz.

Stability under ambient conditions for excitation far above threshold at 400 nm and 4 kHz (translates to four times effective exposure compared with stability reported at 1 kHz laser systems). Perovskites are well known for photo-induced degradation and structural changes on various time scales from ms (single shot) to days. Compared to laser-induced degradation of dye molecules, lead halide perovskites have also been reported to show healing/annealing of photoinduced defects in the dark. For the layered ferroelectric *i*-PCPB perovskite studied herein, we find rather stable PL with negligible annealing, light soaking or degradation effects. Contrary to ordinary PL, superfluorescence first increases as a function of light exposure and degrades after a certain time (turning point approx. 12.000.000 laser shots). The distinct dynamics support the hypothesis, that SF in lead-halide perovskites is driven by a different species of photoexcited states, potentially defect states as a recent study found (*11*). The observed degradation trends did not affect any of the results presented in the main paper because the measurements were completed faster within a few minutes. A notable exception may be the sample-rotation experiment shown in Fig.2 which exposed the same region over an extended period of 90 minutes to the pump laser.

We expect that encapsulation of the samples after preparation in inert atmosphere prevents any water vapor and oxygen exposure and can greatly enhance stability. Likewise, optical excitation closer to the visible (e.g. 480 nm or 515 nm) typically lowers the rate of photodegradation.

## Section 15: Polarization resolved PL and SF microscopy

Microscopy was performed using an Olympus UPlanFLN 40×NA 0.75 objective collecting the PL/SF excited from the backside with an ytterbium-based ultrafast solid-state regenerative amplifier (Pharos, light conversion) and optical parametric amplifier tuned at 400 nm at a repetition rate of 50 kHz. The collimated PL/SF was imaged using a CMOS camera (U3-3880CP Rev.2.2, IDS imaging). Polarization resolution was implemented through a wire-grid polarizer (WP25M-VIS, Thorlabs). In total, 72 images of different analyzer angle were acquired, and, after binning of each 16x16 pixels, the DOP and polarization orientation were calculated using the Malus fit.

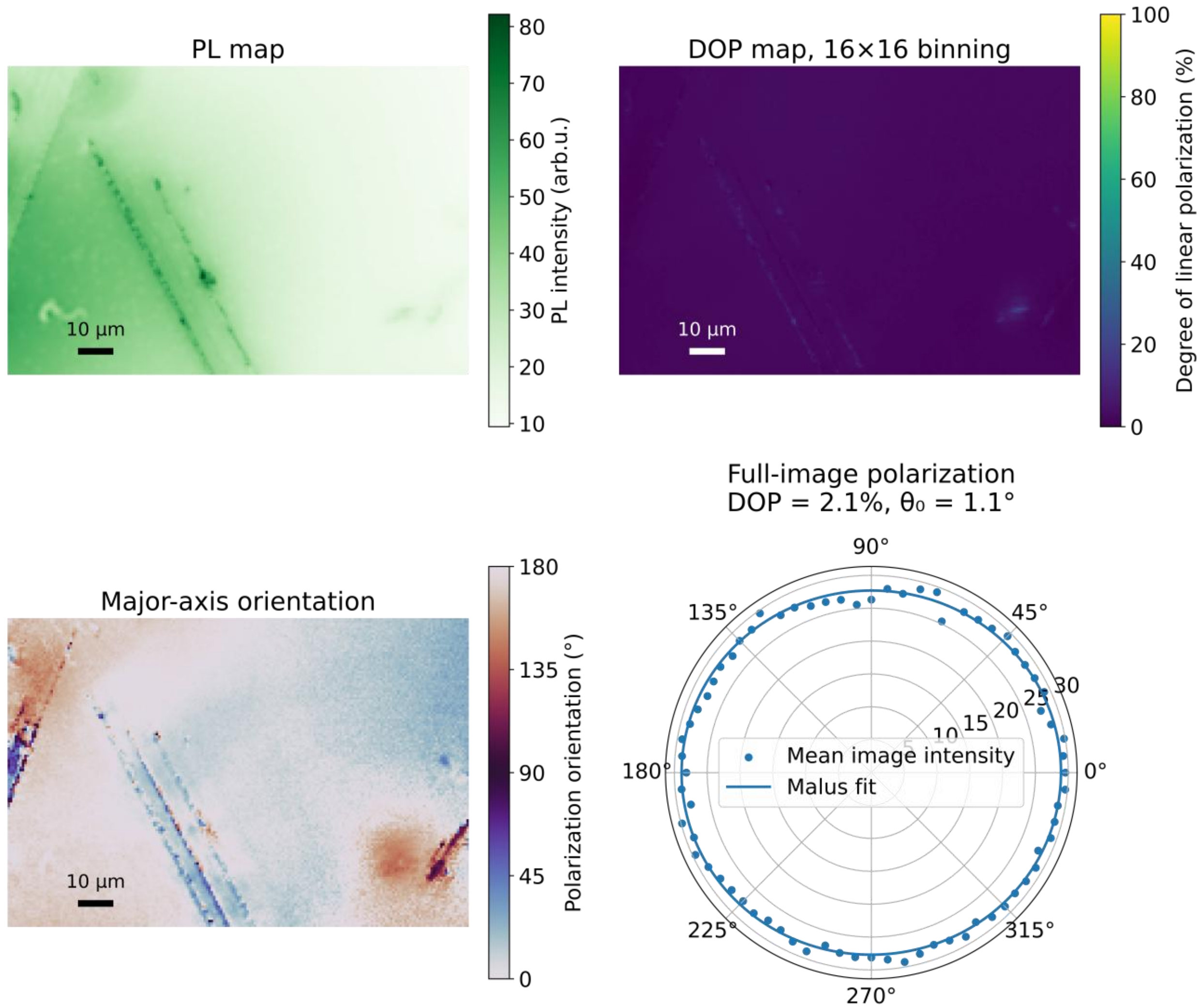

Figure S15: Spatial PL (excitation below SF threshold), DOP, polarization orientation maps and Malus fit of the averaged intensity across the entire image.

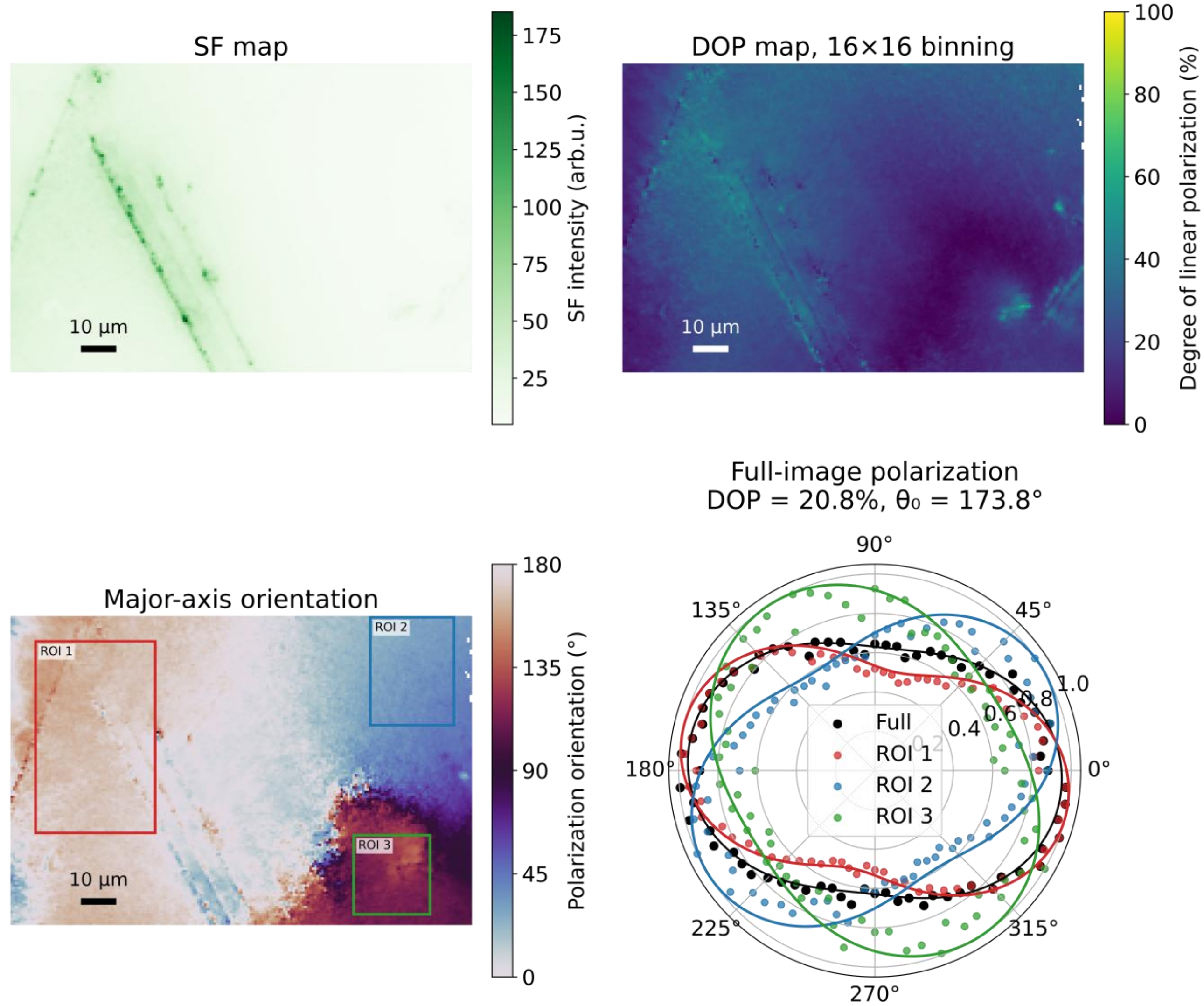

Figure S16: Spatial SF (excitation above SF threshold), DOP, polarization orientation maps and Malus fit of the averaged intensity across the entire image.

Figure S16 shows the same region as Fig. S15 pumped far above threshold such that most of the camera signal stems from SF. Intensity maps in Figs. S16 and S15 cannot be compared quantitatively because exposure time and analog gain of the camera varied between both measurements. Compared to below threshold (Fig. S15), we observe significant degrees of polarization (see also histogram in Fig. S17). The analysis of the major axis orientation reveals several regions with distinct orientation of the linearly polarized SF. This indicates that spatially (and thus temporally) separated regions develop SF independently with the polarization determined by the local anisotropies. The polar plot in Fig. S16 shows that the macroscopic DOP emitted by several regions exceeds the total DOP of the full image because some regions emit orthogonally polarized SF.

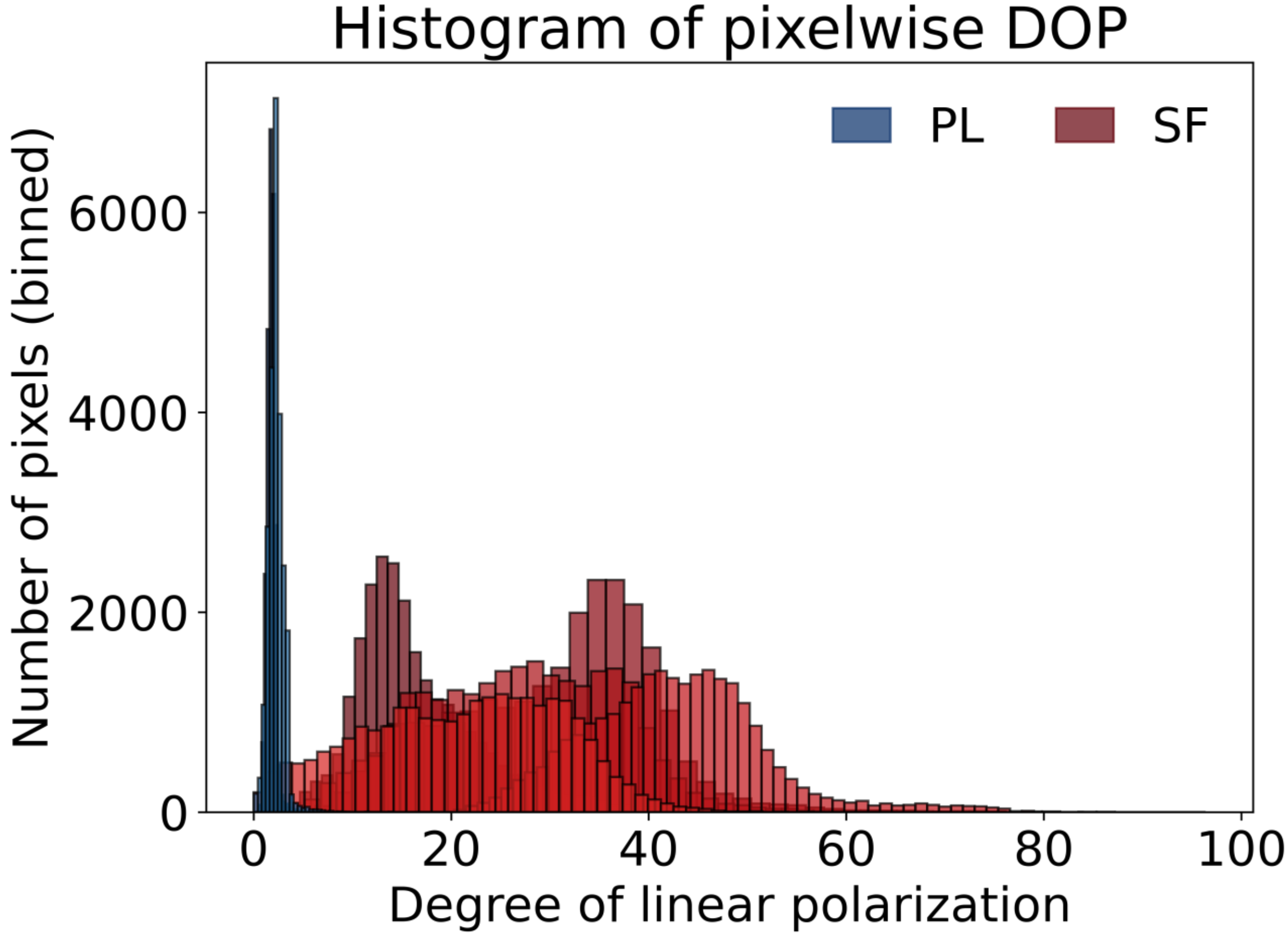

Fig. S17 Degree of polarization of SF and PL for 16x16 binned pixels of multiple regions studied in polarization-resolved microscopy.

The overall distribution confirms that SF shows stronger linear polarization compared to PL and matches the DOPs found for 200 µm sized spots presented in Fig. 3A, indicating that local anisotropies limit the extent of macroscopic SF. Here, without spectra filtering or distinction between PL and SF, there is a cutoff at large DOP for SF, presumably because of unpolarized PL present in the recorded intensity.

**Section 16: Robustness of linearly polarized SF findings**

Linearly polarized SF emission (and negligible linear polarization of PL) was found via several detection and collimation channels.

- time averaged detection with CCD camera and wire-grid polarizer, collimated by lenses of focal distance > 30 mm or high NA objectives

- time-averaged detection using multi-mode optical fiber cables and fiber spectrometers, polarization resolution through wire-grid polarizers, collimated either by off-axis parabolic mirrors or lenses with focal lengths of 30-75 mm.

- up-converted, gated (few 100 fs) detection using off-axis parabolic mirrors for collimation and a spectrometer for the up-converted uv-light

The robustness of our findings related to several different collimation and detection methods rule out most artifacts such as stress birefringence, linear dichroism of optical components or polarization-dependent detection (such as spectrometer gratings or detection). Further, each setup was used without further change to characterize the polarization state of PL and the SF feature of 3D perovskite films which each resulted in negligible degrees of linear polarization.

## Section 17: Characteristic SF scaling at finite temperature:

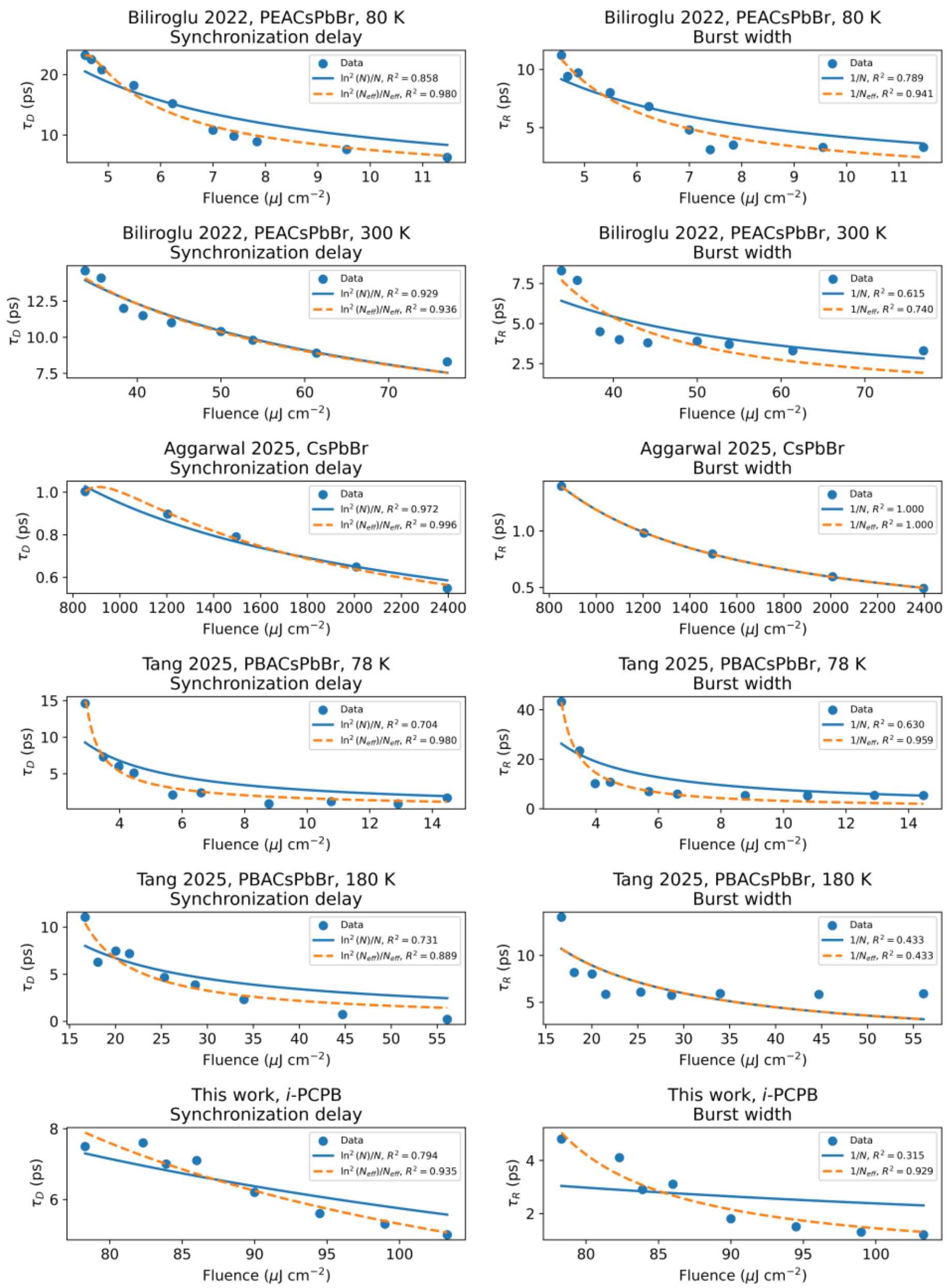

Fig. S18 Synchronization delay and Burst width as reported in the literature with pure Dicke fits (blue, N= A·F) and modified Dicke model (orange, N = A·F – $N_{th}$)

The established key signatures of SF in the literature, distinguishing it from amplified spontaneous emission (ASE), are the delayed emission and the scaling of the delay and pulse width with the number of emitters N. Polder et al. derived these scalings as $\tau_D \propto \frac{\ln^2 N}{N}$ and $\tau_R \propto \frac{1}{N}$ in the absence of any dephasing. Thus, these ideal coherent relations are most likely not applicable to room-temperature SF observed in solids. While all previous studies agree, that N can be tuned with fluence, the exact relationship is ambiguous. A longstanding debate in 2D and quasi-2D lead-halide perovskites is, whether excitons are actually the majority carrier population or if they are only transiently formed from a plasma of unbound electrons and holes. E.g., Findik et al. (*12*) used ideal Dicke-model scaling but with the relation $N = F^2$ due to free carriers forming minority excitons. For simplicity, we will restrict our analysis here to the monomolecular limit, assuming the number of excitons to scale linearly with the fluence. As we demonstrate below, our implementation of the Maxwell-Bloch formalism in the presence of a dephasing term yields the characteristic scalings $I_{peak} \propto N^2$, $\tau_D \propto \ln^2(N)/N$ and $\tau_R \propto 1/N$ only if a threshold exciton density $N_{th}$ is considered. In Fig. S18, we applied these modified scalings to previously reported experimental data and our data and obtain a better fit for all quasi-2D datasets (*2*, *13*, *14*). Only the data for 3D perovskite SF, obtained at fluences 20x larger, by Aggarwal et al. (*13*) are equally well described by both models.

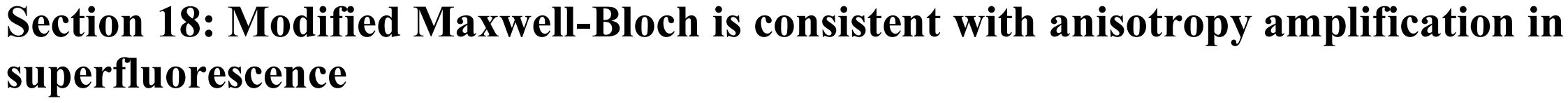

**Section 18: Modified Maxwell-Bloch is consistent with anisotropy amplification in superfluorescence**

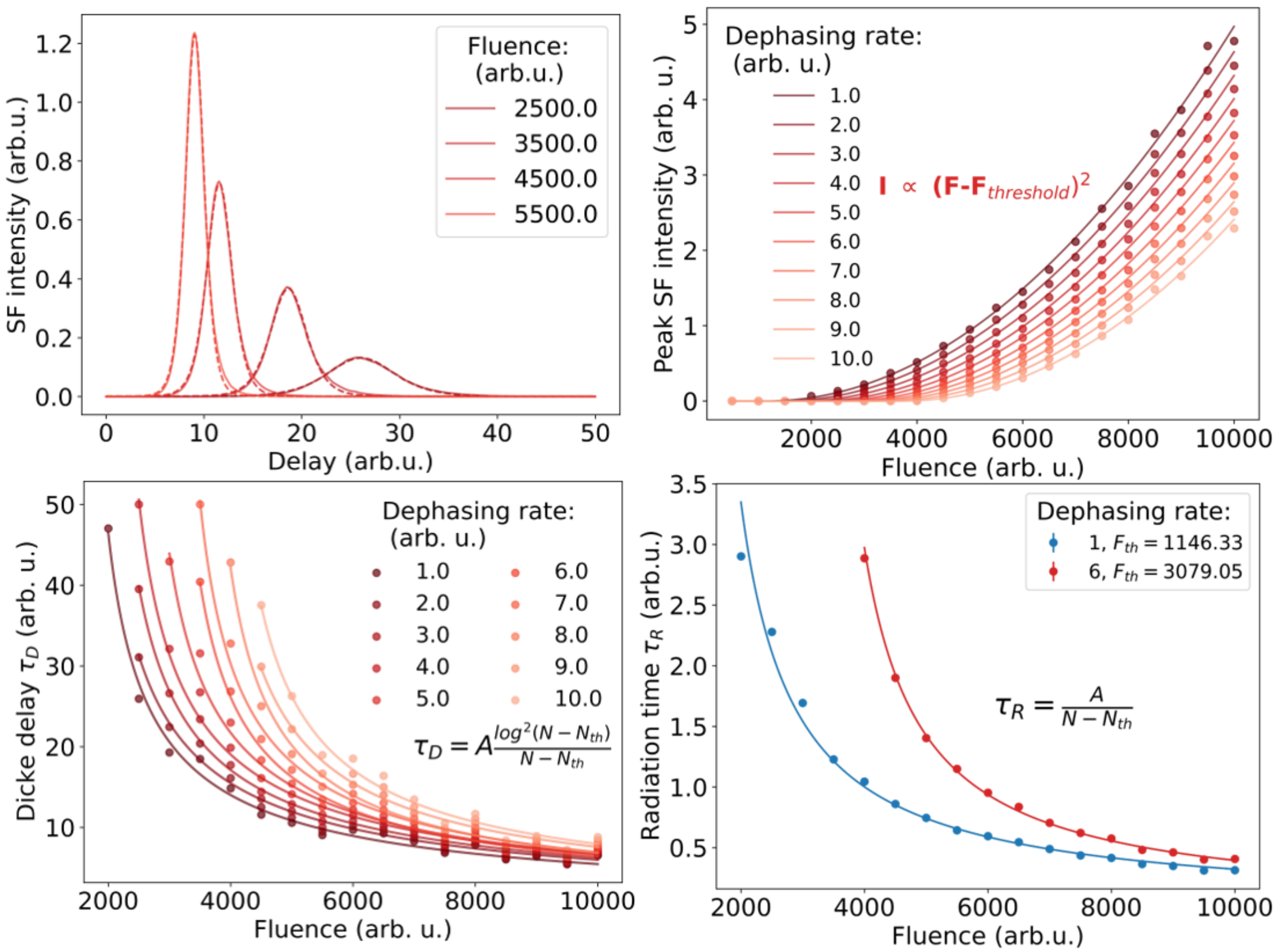

Figure S19. Simulated superfluorescence characteristics and their scaling with the number of emitters, N, using our semiclassical mean-field Maxwell–Bloch model including finite dephasing. In the simulations shown here, N is assumed to scale linearly with excitation fluence. The model reproduces characteristic signatures of room-temperature superfluorescence, including the nonlinear increase of the burst peak intensity, shortening of the radiation time, and reduction of the emission delay with increasing N, albeit ideal Dicke-model relationships are obtained for $N_{eff} = N-N_{th}$.

The emitted intensity is calculated from the squared magnitude of the macroscopic optical polarization, $I(t) \propto |P(t)|^2$. The coupling of each transition dipole to the common radiation field depends on its orientation through the corresponding scalar product. Consequently, dipoles aligned more closely with a given field polarization contribute more strongly to the collective polarization associated with that mode.

Importantly, even for κ=0, the single-mode mean-field model generates strongly polarized individual SF bursts, but with a polarization direction that is random from burst to burst. This differs from our single-shot measurements of the three-dimensional mixed-cation perovskite, for which the SF emission remains weakly polarized even at the level of individual bursts (see Fig. S8). The orientation-dependent single-mode coupling used in the present model therefore does not appear to describe the cubic lead-bromide control sample adequately.

We propose that the reduced structural symmetry and anisotropic excitonic transition dipoles of the two-dimensional ferroelectric perovskite *i*-PCPB make its cooperative light–matter coupling more sensitive to weak local orientational biases. Such biases could originate, for example, from directional crystallization or solvent-flow fronts during spin coating. The collective buildup of macroscopic polarization can then convert a weak microscopic anisotropy into strongly linearly polarized SF emission.

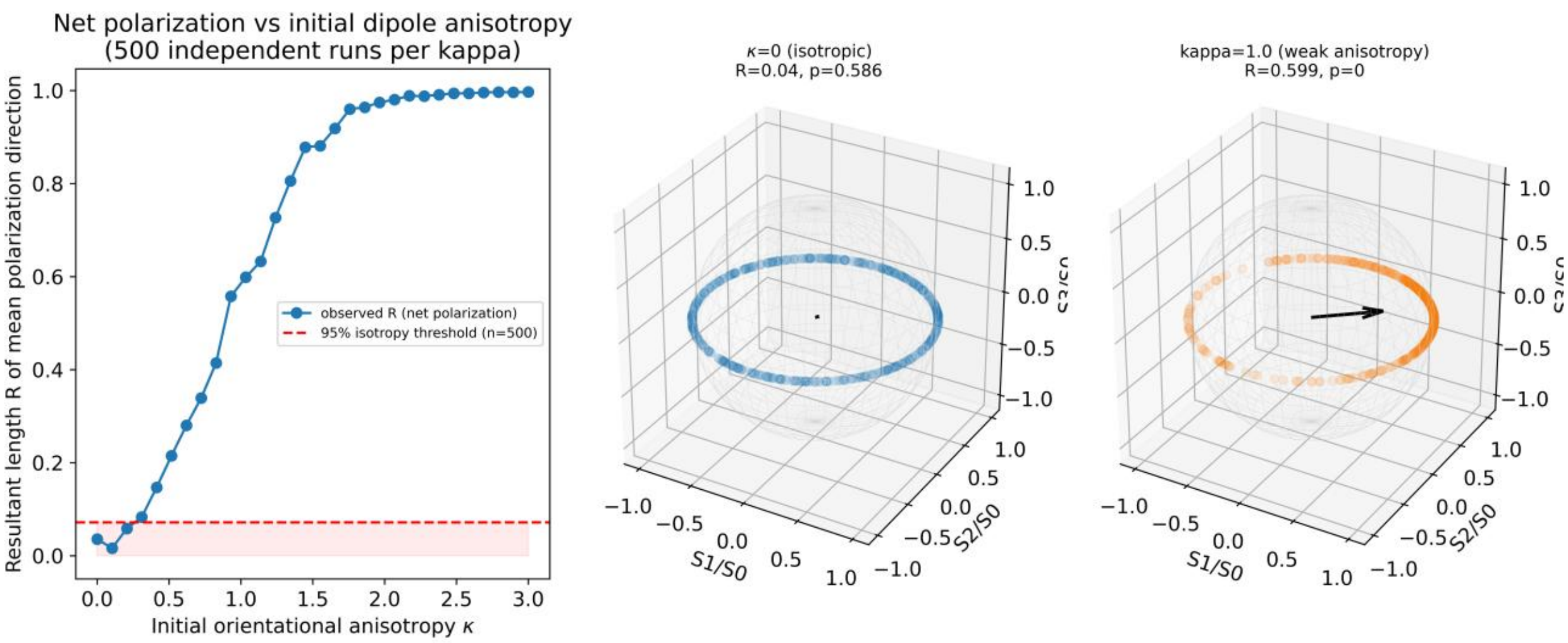

Figure S20. Mean and distributions of the polarization states of individual simulated SF bursts for different values of the anisotropy parameter κ. The parameter κ controls the three-dimensional orientational distribution of the transition dipoles, with κ=0 corresponding to an isotropic distribution and increasing κ representing progressively stronger alignment about a preferred axis. For isotropically distributed dipoles, the ensemble-averaged Stokes vector obtained from 500 independent bursts approaches zero, although each individual simulated burst is strongly polarized and its polarization orientation varies randomly between bursts. Introducing even a weak orientational bias, for example κ=1, produces a clearly anisotropic distribution of the burst Stokes vectors.

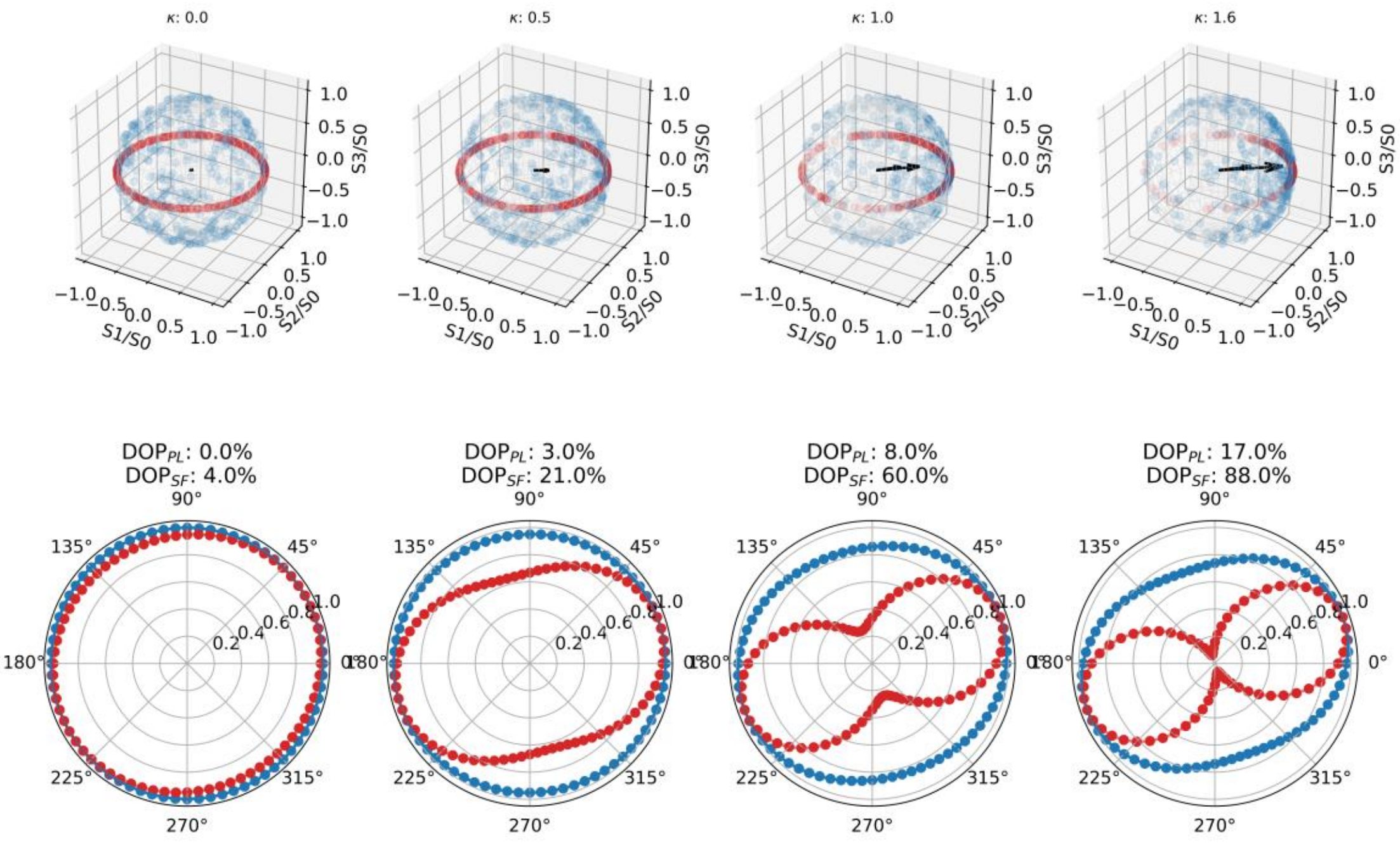

Figure S21. Comparison of the expected degree of linear polarization for incoherent spontaneous emission and for SF as a function of κ. For spontaneous emission, the intensities of the individual dipoles are added incoherently. In the SF simulation, by contrast, the dipoles contribute to a common macroscopic polarization, and their orientation-dependent coupling promotes collective emission along the favored polarization direction. The model therefore amplifies weak microscopic orientational anisotropies. In the parameter range considered here, dipole distributions that produce only approximately 3–8% linear polarization in spontaneous emission result in simulated SF degrees of polarization of approximately 20–60%.

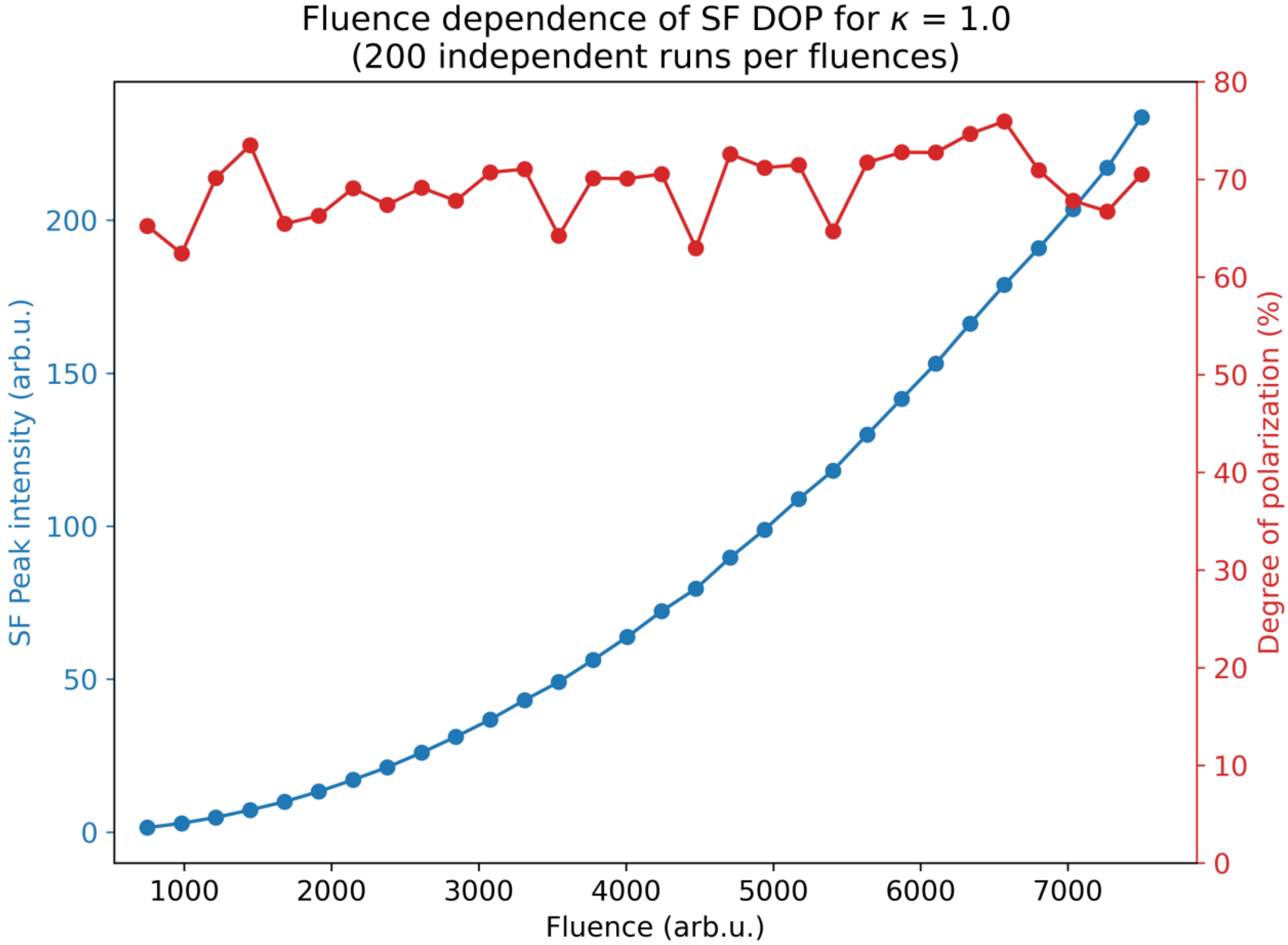

Fig. S22: Simulated data for κ=1 matches fluence-independent DOP as observed experimentally in Fig. 2E.